\documentclass[aps,superscriptaddress,amsmath,amssymb,twocolumn,amsfonts,floatfix,nofootinbib,longbibliography]{revtex4-1}
\usepackage{graphicx}
\usepackage{color}
\graphicspath{{Pictures/}}
\usepackage{hyperref}
\usepackage{soul}
\usepackage{natbib}
\hypersetup{pdfnewwindow=true, colorlinks=true, linkcolor=ForestGreen, anchorcolor=blue, citecolor=Purple, filecolor=blue, menucolor=blue, urlcolor=magenta}
\usepackage{cleveref}
\usepackage{color}
\usepackage{float}
\usepackage{soul}
\usepackage{braket}
\usepackage{enumitem}
\usepackage{marginnote}
\usepackage[normalem]{ulem}

\usepackage{bm}
\usepackage{MnSymbol}
\usepackage{titlesec}

\usepackage{amsmath}
\usepackage{mathtools}
\usepackage{braket}
\usepackage{graphicx}
\usepackage{amssymb}
\usepackage[dvipsnames]{xcolor}
\usepackage{multirow}
\usepackage{xfrac}
\usepackage{textcomp}
\usepackage{float}
\usepackage{enumerate}
\usepackage{subfigure}
\usepackage{stackengine}
\usepackage{booktabs}
\usepackage{physics}

\mathchardef\mhyphen="2D 

\newcommand\bea{\begin{eqnarray}}
	\newcommand\eea{\end{eqnarray}}
\newcommand\beq{\begin{equation}}  
	\newcommand\eeq{\end{equation}}

\usepackage{sidecap,tikz}

\begin{document}
	
	
	\title{Superconducting diode effect via Floquet topological Fulde-Ferrell phase in driven Rashba nanowire}
	
	\author{Sayak Bhowmik}
	\email{sayak.bhowmik@iopb.res.in}
	\affiliation{Institute of Physics, Sachivalaya Marg, Bhubaneswar, Orissa 751005, India}
	\affiliation{Homi Bhabha National Institute, Training School Complex, Anushakti Nagar, Mumbai 400094, India}
	
	\author{Arijit Saha}
	\email{arijit@iopb.res.in}
	\affiliation{Institute of Physics, Sachivalaya Marg, Bhubaneswar, Orissa 751005, India}
	\affiliation{Homi Bhabha National Institute, Training School Complex, Anushakti Nagar, Mumbai 400094, India}
	
	\author{Tanay Nag}
	\email{tanay.nag@hyderabad.bits-pilani.ac.in}
	\affiliation{Department of Physics, BITS Pilani-Hyderabad Campus, Telangana 500078, India}

	\begin{abstract}
		
		Much has been studied on Floquet engineering in Rashba nanowire model regarding topological superconductivity hosting Majorana $0$- and anomalous $\pi$-modes, 
		here we theoretically investigate the possible emergence of finite momentum Fulde-Ferrell (FF) superconducting state in quasi-energy of the above model under 
		the periodic modulation of in-plane and out-of-plane magnetic fields while the static limit does not host a FF ground state. We demonstrate controllable switching between Floquet Majorana $0$- and $\pi$-modes via reversal of the supercurrent direction, revealing pronounced nonreciprocal supercurrent signatures which is a manifestation of the FF pairing. We validate the onset of FF pairing following a self-consistent mean-field analysis where externally applied supercurrent facilitates nonreciprocal signatures in quasi-energy spectra. The above findings directly indicates to the intriguing phenomenon of superconducting diode effect (SDE). The drive amplitude serves as a parameter to regulate the diode effciency with chemical potential. Our study thus reveals a rich interplay between Floquet topological superconductivity and finite-momentum FF pairing, providing a tunable way to switch between different Floquet Majorana modes and realize the SDE with high efficiency.

	\end{abstract}
	
	\maketitle
	
	{\textcolor{blue}{\textit{Introduction}}}- In recent times, topological superconductivity (TSC) has emerged as one of the central themes in modern quantum condensed matter physics owing to its ability to host Majorana zero modes (MZMs). These charge-neutral zero-energy quasiparticle excitations have attracted enormous attention due to their remarkable non-Abelian braiding statistics and their potential applications in fault-tolerant topological quantum computation ~\cite{Ivanov,KITAEV20032,Stern2010,CNayak}. Following Kitaev's pioneering theoretical proposal of a one-dimensional (1D) spinless $p$-wave superconductor~\cite{Kitaev2009,Kitaev_2001}, several theoretical proposals and experimental realizations have emerged to engineer Kitaev's physics regarding MZMs~\cite{Leijnse_2012,Aguado2017,Alicea_2012,Lutchyn_Sau,Mourik,Zhang_Qi,Alicea_2012,Yuval_Oreg_Oppen,Beenakker}. Among them, Rashba semiconducting 1D nanowires proximitized to conventional $s$-wave superconductors under external Zeeman fields emerge as one of the most promising and extensively explored hybrid platforms~\cite{Lutchyn_Sau, Mourik,SAU_PRL, PRL_SAU,Rainis,Beenakker}.
	
	In recent literature, the exploration of topological matter has been extended to periodically driven quantum systems through the sophisticated framework of Floquet engineering, facilitating the advent of exotic dynamical topologically non-trivial phases beyond their static counterparts~\cite{OKA2009,lindner2011floquet,Barnea,Mondal_1,Mondal2,Eckardt,ghosh2024generation,Ghosh2022,Rudner2013,Yao2017,OKA}. In particular, driven superconducting systems can host not only the conventional Floquet Majorana $0$-modes, but also robust finite energy $\pi$-Floquet Majorana modes ($\pi$-FMMs), representing intrinsically dynamical topological excitations with no static analog.  
	
	Moreover, the interplay of spin-orbit coupling (SOC), superconductivity, and Zeeman fields in such systems, simultaneously breaking time-reversal and inversion symmetries, not only promotes the emergence of topologically nontrivial MZMs, but also provides an intriguing platform for realizing unconventional finite-momentum superconducting phases such as the Fulde-Ferrell (FF) state~\cite{Fulde_1964,Larkin_1964,Sayak_2025}. Such finite-momentum paired phases 
	are generically probed by externally injected supercurrent, constituting one of the key ingredients behind the 
	nonreciprocal charge transport in superconductors, thereby giving rise to the intriguing phenomenon of the superconducting diode effect (SDE)~\cite{Yanasediode,Yanaseprl,nadeem2023superconducting, LiangPNAS,SLlicprl,Yanasediode,Picoli,Nagaosanjp,Qu2013natcom,Lossdiode,ruthvik2025field,pal2025topological}. 
	The time-reversal and inversion symmetries are found to be broken to generate finite 
	SDE~\cite{nadeem2023superconducting,PhysRevLett.87.236602,PhysRevLett.130.136301,PhysRevB.99.245153,ando2020observation,Lossdiode,PhysRevB.109.094501}. Given the background of Floquet dynamics and SDE, the Floquet generation of FF states and subsequently appearance of SDE are unexplored as far as the interplay between supercurrent, FMMs and non-reciprocity of critical current are concerned.
	
	In light of these developments, in this Letter, we investigate the possible emergence of FF superconducting state, considering a Rashba nanowire proximitized to a bulk $s$-wave superconductor, under the periodic modulation of both in-plane and out-of-plane magnetic fields while the underlying static limit does not support any FF ground state. The injection of supercurrent is essential to yield SDE in the system, hence, allowing us to achieve a controllable topological transitions between $0$- and $\pi$- FMMs by varying the injection current. Notably, our comprehensive analysis reveals that oppositely directed supercurrents selectively lead to the emergence of Floquet TSC phases hosted by \(0\)- and \(\pi\)-FMMs, demonstrating a supercurrent-driven topological transition between distinct FMMs. We further validate our work through the self-consistent mean-field analysis where we find that the periodic drive stabilizes a Floquet-induced 
	FF superconducting state, giving rise to pronounced nonreciprocal supercurrent response intertwined with Floquet topology. The resulting asymmetry of the critical supercurrents leads to SDE with a significant diode efficiency upon varying the drive parameters.

	\begin{figure}[t!]
		\centering \includegraphics[width=\columnwidth]{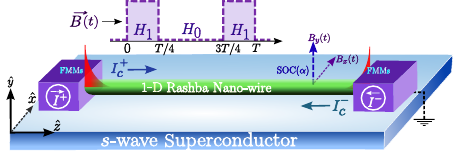}
		\caption{
			We illustrate the schematic of a 1D Rashba nanowire (green), featuring Rashba SOC of strength $\alpha$, that is placed in close proximity to a bulk $s$-wave superconductor (blue) in presence of external Floquet step drive in terms of the magnetic field $\mathbf{B(t)}=(B_x(t), B_y(t))$. This setup realizes  topological FMMs (red spikes) localized at the end of the NW, accompanied by non reciprocal critical supercurrents $I^+_c \ne I^-_c$ highlighting Floquet induced SDE.}
		\label{fig:schematic}
	\end{figure}
	{\textcolor{blue}{\textit{Model}}}- We begin by considering a 1D semiconducting nanowire (NW) featuring Rashba SOC placed in close proximity to a conventional $s$-wave superconductor. To stabilize the superconducting order in our system, we consider a local attractive Hubbard interaction $\mathcal{H}_I=-\frac{U}{2}\sum_{s^\prime} \int d^{3} {\bf{r}} c_{s}^\dagger({\bf{r}})c_{s^\prime}^\dagger({\bf{r}})c_{s^\prime}({\bf{r}})c_{s}({\bf{r}})$ within the bulk $s$-wave superconductor followed by a mean-field decoupling in the $s$-wave channel: $\Delta(r) = -U \langle c_{r\downarrow} \, c_{r\uparrow} \rangle$, where $U$ is the Hubbard interaction strength and $s$ denotes electronic spin $\{\uparrow,\downarrow\}$. In presence of externally injected supercurrent, realized through finite Cooper pair momentum $q$, the superconducting order parameter takes the form $\Delta(r)= \Delta e^{iqr}$. We assume this superconducting correlations are uniformly induced in the nanowire via proximity effect while adopting a mean-field superconducting order parameter. This leads to the following tight-binding lattice Hamiltonian in the BdG basis as
	\begin{align}
		H_0 ={}& - t \sum_{n,s} c_{n+1,s}^\dagger c_{n,s}
		- \sum_n \Delta e^{iqn}
		\left(c_{n,\uparrow}^\dagger c_{n,\downarrow}^\dagger + \mathrm{H.c.}\right) \nonumber \\
		&+ i\alpha \sum_{n,s,s'} 
		c_{n+1,s}^\dagger (\sigma^y)_{ss'} c_{n,s'} + \mathrm{H.c.}
		- \mu \sum_{n,s} c_{n,s}^\dagger c_{n,s}\ ,
		\label{lattice}
	\end{align}
	where, $t$ is the bare electron hopping strength, $n$ denotes the lattice site index, $\mu$ denotes the chemical potential and $\alpha$ represents the strength of Rashba SOC manifested by a spin-dependent nearest-neighbor hopping with imaginary amplitude. 

\begin{figure*}[ht]
	\centering
	\includegraphics[width=1.0\linewidth]{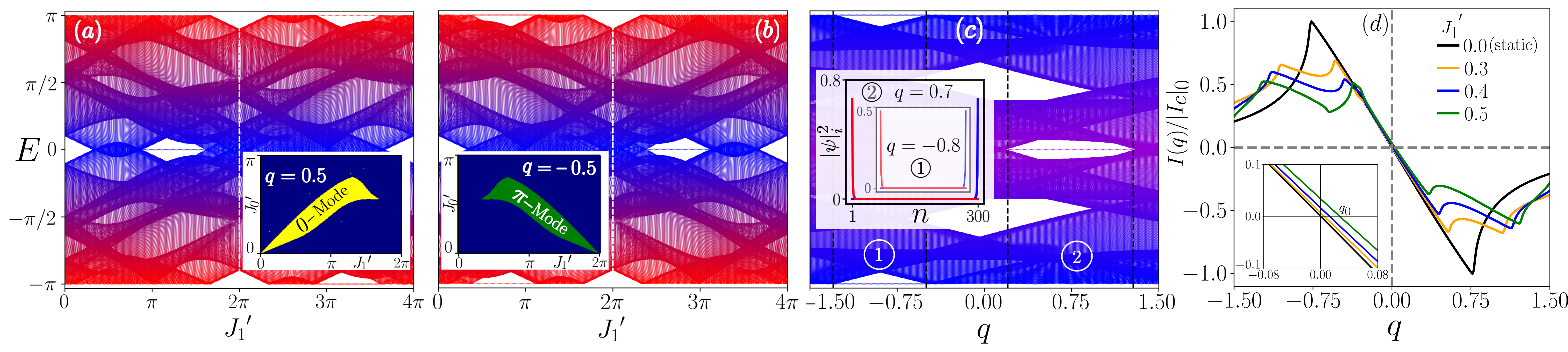}
	\caption{
		We depict the spectral properties of the driven Rashba NW model by demonstrating the Floquet quasi-energy spectrum $E_n$, computed from Eq.~(\ref{eq:FloquetU}), as a function of $J^{\prime}_1$ in panels (a,b) for $q=0.5$ and $q=-0.5$, respectively, with $B_y=0.6$ and $J^{\prime}_0=0.6\pi$. The in-gap states (white region) correspond to the emergence of topological $0$- and $\pi$-FMMs. Insets in panels (a) and (b) exhibit the corresponding topological phase diagrams in the $J^{\prime}_1$-$J^{\prime}_0$ plane, where the yellow (green) region denotes Floquet TSC phases with $0$- ($\pi$-) FMMs yielding $P_x=0.5$, while the blue region represents the trivial phase with $P_x=0$. Panel (c) shows $E_n$ as a function of the Cooper-pair momentum $q$ for $(J^{\prime}_1$, $J^{\prime}_0)=(0.8\pi, 0.46\pi)$, highlighting the emergence of topological $\pi$- ($0$-) modes in the regions encircled $1$ ($2$). The insets display the site-resolved probability amplitude $|\psi_i|^2$ for representative states in region $1$ ($2$) for $q=-0.8$ ($q=0.7$) asscociated with $\pi$- ($0$-) FMMs. The supercurrent density as a function of $q$ is shown in panel (d) for different driving strength $J_1^{'}$ with $J_0^{'}=0.6\pi$. 
		The finite shift of FF ground state Cooper pair momentum satisfying $I(q_0)=0$ is shown in the inset of panel (d). The other model parameters are considered 
		as follows: $(\mu, \alpha, t,\Delta)=1.0J_0$.}
	\label{fig:Fig1}
\end{figure*}	

{\textcolor{blue}{\textit{Floquet drive protocol}}}- In our analysis, the above static model (Eq.~(\ref{lattice})) is subjected to an externally applied  time-periodic Zeeman fields $\mathbf{B}(t)=(B_x(t),B_y(t))$, implemented via three step driving protocol (see schematic in Fig.~\ref{fig:schematic}). 
The time-dependent Hamiltonian is given by
\begin{equation}
	H_{\mathrm{step}}(t)=
	\begin{cases}
		J_1 H_1, & t \in [0, T/4)\ , \\
		J_0 H_0, & t \in [T/4, 3T/4)\ , \\
		J_1 H_1, & t \in [3T/4, T]\ ,
	\end{cases}
	\label{drive}
\end{equation}
where, $ H_1 =\sum_{n,s,s'}  c_{n,s}^\dagger \, (\mathbf{B} \cdot \boldsymbol{\sigma})_{ss'} \, c_{n,s'}$ with $T$ being the time period of the drive. All the model parameters in the Hamiltonian $H_0$ and $H_1$ are expressed in energy units $(J_0$, $J_1)$ respectively, with $|\mathbf B|=1$ for simplicity, throughout our analysis. Following the standard Floquet formalism to capture the stroboscopic dynamics, we construct the Floquet operator, given as
\begin{align}
	U(T) &= \mathcal{T} \exp\left(-i \int_0^T dt\, H_{\mathrm{step}}(t)\right)=\exp\left(-i\mathcal{H}_{F} T \right)  \nonumber \\
	&= e^{-i J_1 H_1 T/4} \, e^{-i J_0 H_0(k) T/2} \, e^{-i J_1 H_1 T/4}\ .
	\label{eq:FloquetU}
\end{align}
Here, $\mathcal{T}$ is the time-ordering operator and $\mathcal{H}_{F}$ denotes the time independent Floquet Hamiltonian. 
	Note that, the FF  ground state in a $s$-wave superconductor is favoured in the presence of static in-plane as well as out of plane magnetic fields 
	and Rashba SOC~\cite{Lossdiode,qu2013topological,hu2019topological}. We diagonalize $U(T)$ to obtain the Floquet quasi-energy spectrum by utilizing the spectral properties of $U(T)$ such that $U(T)\, |\psi_n\rangle = e^{-i E_n T}\, |\psi_n\rangle$, where $E_n$ represents the quasi-energy corresponding to the Floquet eigenstate $|\psi_n\rangle$. We further define dimensionless parameters: $(J^{\prime}_1, J^{\prime}_0)=(J_1 T,J_0 T)$ such that the quasi energy spectrum lies within the first Floquet Brillouin zone, $E_n \in[-\pi,\pi]$. 
	
	Having outlined the Floquet machinery, we now compute the supercurrent flowing across the bond $\langle i,j\rangle$ employing periodic boundar condition (PBC), given as~\cite{Zhu_2016}
	\begin{align}
		\mathcal{I}_{\langle i,j\rangle} 
		= \frac{2e}{i\hbar} \sum_{s,s'} \sum_m 
		\Big[ \,
		- t'_{ji}(s,s')\, u^{m*}_{j,s} u^{m}_{i,s'} n_F(E_m) \nonumber \\
		\qquad\quad
		+ t'^*_{ji}(s,s')\, v^{m}_{j,s} v^{m*}_{i,s'} \big(1 - n_F(E_m)\big)
		- \mathrm{c.c.} \, \Big] \ ,
		\label{eqn:current}
	\end{align}
	where, $t'_{ji}$ encodes the effective hopping on the bond $\langle i,j\rangle$, incorporating both the bare hopping amplitude $t$ and SOC $\alpha$, as discussed in Eq.~(\ref{lattice}). The function $n_F(E_m)$ denotes the Fermi-Dirac distribution. The quantities $u^m_{i,s}$ and $v^m_{i,s}$ represent the standard particle and hole components of the $m^{\rm{th}}$ Floquet eigenstate $|\psi_m\rangle$, respectively. The net supercurrent density $I(q)$  can be obtained by averaging $\mathcal{I}_{\langle i,j\rangle}$ over all the bonds ${\langle i,j\rangle}$.
	
	{\textcolor{blue}{\textit{Spectral analysis and appearance of FMMs-}}} 
In order to illustrate the emergence of topological $0$-and $\pi$-FMMs, we obtain the Floquet Hamiltonian $\mathcal{H}_{F}$ by employing open boundary condition (OBC) and numerically compute the quasi-energy spectrum $E_n$. Importantly, for the chosen driving protocol in Eq.~(\ref{drive}), the quasienergy spectrum satisfies a generic relation: $E_n(J^{\prime}_1+2\pi)=E_n(J^{\prime}_1)+\pi$ (see Supplemental material \cite{supp} for details). Notably, upon introducing a supercurrent via a finite Cooper-pair momentum $q$, the spectrum obeys
\begin{equation}
E_n(J^{\prime}_1,-q)=E_n(2\pi-J^{\prime}_1,q)+\pi \ .
\label{switch}
\end{equation}
The $\pi$-shift in the quasienergy spectrum, controlled by the externally tunable Floquet driving strength $J_1'$ and finite Cooper-pair momentum mediated supercurrent, guarantees the emergence of dynamical topological $\pi$-FMM in the complementary parameter regime restricting the simultaneous presence of $0$- 
and $\pi$-FMMs.

As a result, the independent control of these parameters following Eq.~(\ref{switch}) enables the switching between topological $0$- and $\pi$-FMMs. Importantly, $0$- and $\pi$-modes are selectively realized in positive and negative values of $q$, respectively reflecting the nonreciprocal nature of the supercurrent. 
To validate the above we numerically compute the BdG quasienergy spectrum $E_n$ by diagonalizing the Floquet Hamiltonian $\mathcal{H}_{F}$ under OBC. The resulting spectra is shown in Fig.~\ref{fig:Fig1} as a function of $J^\prime_1$ for $q=0.5$ (see Fig.~\ref{fig:Fig1}(a)), and $q=-0.5$, (see Fig.~\ref{fig:Fig1}(b)). A direct comparison between Figs.~\ref{fig:Fig1}(a) and (b) further reveals the fact that upon reversal of $q \rightarrow -q$, 
the quasienergy spectrum follows the mapping $J^{\prime}_1 \rightarrow 2\pi - J^{\prime}_1$ with a $\pi$ energy shift, in agreement with Eq.~(\ref{switch}) discussed above. We further present the quasienergy spectrum as a function of $q$, revealing current-induced topological phase transitions as shown in Fig.~\ref{fig:Fig1}(c). Topological Floquet modes emerge only beyond a finite $q$, indicating a supercurrent-driven transition. The positive- and negative-$q$ sectors display complementary behavior, with $0$- and $\pi$-FMMs appearing in opposite sectors, consistent with the $\pi$-shift relation. This establishes current-driven switching between $0$ and $\pi$-FMMs.


\begin{figure*}[t!]
\centering
\includegraphics[width=1\linewidth]{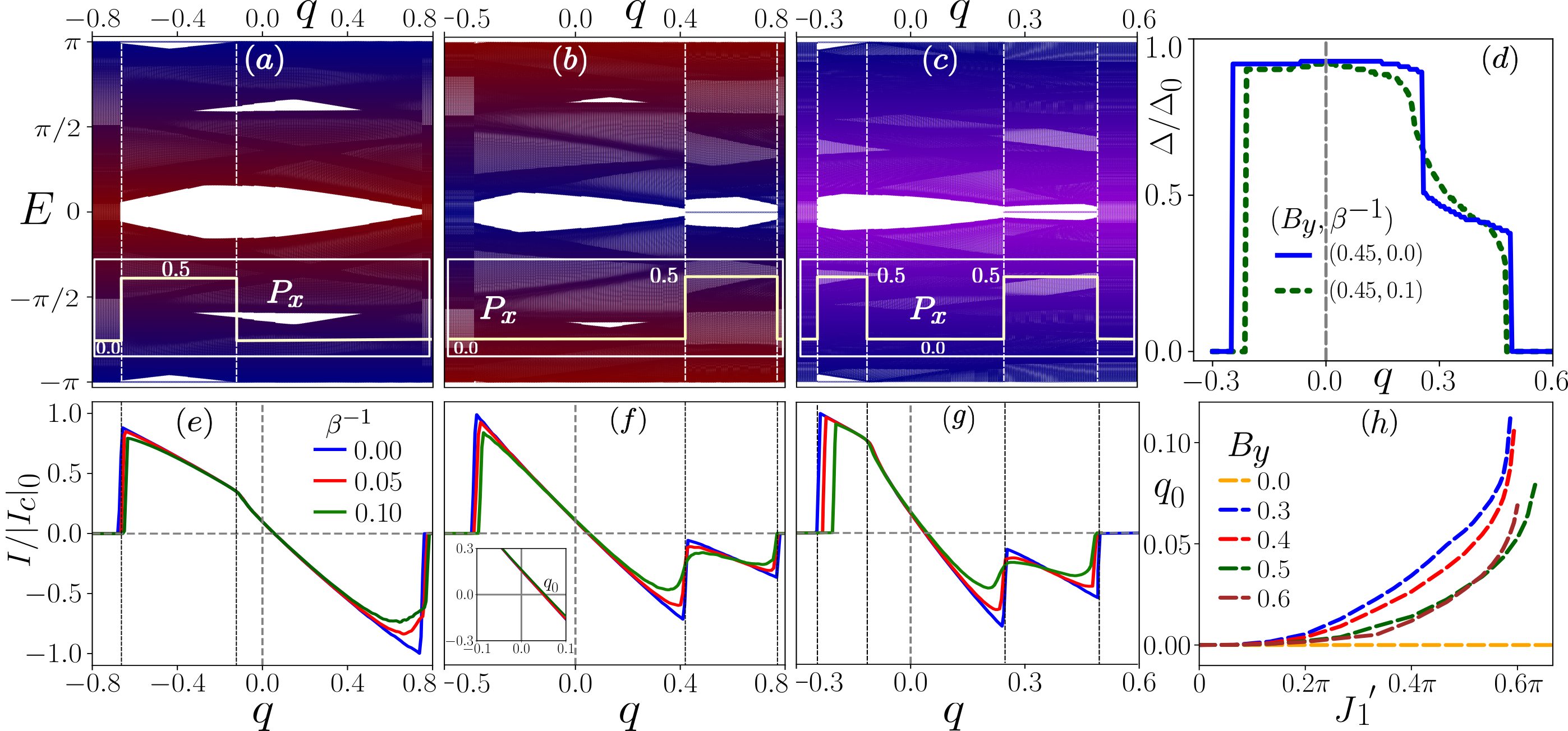}
\caption{
We demonstrate the Floquet quasienergy spectrum, along with the topological invariant $P_x$, obtained within the self-consistent framework exploting 
Eq.~(\ref{eqn:free_energy}), as a function of the finite Cooper-pair momentum $q$ in panels (a,b,c). Panel (a) depicts the emergence of topological $\pi$-FMMs exclusively in the negative-$q$ sector, panel (b) features $0$-FMMs in the positive-$q$ sector, while panel (c) simultaneously hosts $\pi$-FMMs in the negative-$q$ sector and $0$-FMMs in the positive-$q$ sector, highlighting the switching between dynamical topological FMMs in FF phase. The parameter set $({J'_1}, {J'_0}, \mu/{J'_0}, B_y)$ is chosen as $(0.5\pi, 0.46\pi, 1.2, 0.45)$ in panel (a), $(0.6\pi, 0.46\pi, 1.2, 0.5)$ in panel (b), and $(0.5\pi, 0.63\pi, 1.5, 0.45)$ in panel (c). The supercurrent density profiles are displayed as a function of $q$ in panels (e,f,g), considering the same set of parameter values as mentioned for panels (a,b,c), respectively, for different temperatures. The topological boundaries hosting FMMs are marked by vertical dashed lines in panels (a,b,c,e,f,g). Panel (d) demonstrates the profile of superconducting order parameter $\Delta(q)$ for different values of $B_y, \beta^{-1}$ with system parameters chosen as mentioned for panel (c). The FF ground state Cooper-pair momentum $q_{0}$ is displayed as a function of ${J'_1}$ for different values of $B_y$ in panel (h) with corresponding parameters chosen as mentioned for panels (b,f). Other model parameters are chosen as $(t, \alpha)= 1{J_0}^{'}, U=2.3t$ along with a finite system of $300$ lattice sites.}       
\label{fig:Fig2}
\end{figure*}	

To appropriately characterize the topological phases and establish the bulk boundary correspondence, we use the bulk topological invariant namely, polarization 
$P_x$~\cite{Resta,Wheeler,KangPRB}, defined as 
\begin{equation}
P_x = \frac{1}{2\pi}\mathrm{Im}[\mathrm{Tr}\{\ln(\mathcal{U}^\dagger W \mathcal{U})\}]\ , \\
\label{polarization} 
\end{equation}  
where, the operator $W = e^{i2\pi \hat{p}}$ with $\hat{p}=\hat{x}/L$ encodes the many-body position operator for a system with length $L$. The matrix $\mathcal {U}$ $(N \cross N_{oc})$ is formulated by column-wise representing the occupied eigen-states of the Floquet operator $U(T)$ (Eq.~(\ref{eq:FloquetU})), employing PBC. The topological phase is characterized by a quantized value $P_x=0.5$, while $P_x=0$ corresponds to the trivial regime. We examine $P_x$ in the $J^{\prime}_1$-$J^{\prime}_0$ plane for $q=0.5$ and $q=-0.5$, as shown in the insets of Figs.~\ref{fig:Fig1}(a) and (b), respectively. The regions with $P_x=0.5$ identify the topological phases hosting $0$-FMMs (yellow regime) and $\pi$-FMMs (green regime) for $q=0.5$ and $q=-0.5$ respectively, thereby substantiating the spectral features discussed above. Notably, a single invariant suffices to characterize the $0$- and $\pi$-modes unanimously in the present context. The invariant consistently identifies the regions hosting $0$- and $\pi$-FMMs, in agreement with the quasi-energy spectra. Notably, $0$- and $\pi$-FMMs never 
appear simultaneously for the parameter regime with a given value of $q$, as evident from the phase diagrams.

All the above mentioned results are obtained within a non-self-consistent framework, assuming a constant superconducting order parameter in order to capture the essential quasienergy structure. The presence of finite $B_y$ in the Floquet drive leads to an asymmetric quasienergy spectrum in terms of $q$ as shown in Fig.~\ref{fig:Fig1}(c). This signals towards generation of nonreciprocal supercurrent and finite-$q$ FF type pairing. Therefore, our analysis clearly indicates an onset of finite Cooper-pair momentum as $I(q=q_0)=0$ for $q_0\ne 0$, ensuring the emergence of FF gound state in Floquet spectrum, in contrast to the static case where $q_0=0$ (see Fig.~\ref{fig:Fig1}(d)). While our analysis provides key spectral insight and a qualitative indication of emerged FF pairing, the superconductivity can be significantly suppressed under a strong periodic drive. We therefore perform a fully self-consistent analysis, discussed next, to  address the key insights about current-induced topological phase transitions, validating the Floquet-induced FF pairing and legitimize the supercurrent non-reciprocity.

{\textcolor{blue}{\textit{Self-consistent mean-field treatment-}}} Following the self-consistent BdG mean-field framework, we determine the superconducting 
order parameter $\Delta(q)$ in the presence of external supercurrent that can be realized via finite Cooper pair momentum $q$. We minimize the condensation energy $\Omega(\Delta,q)=F(\Delta,q)-F(0,q)$ with respect to $\Delta$ for a given $q$ to obtain the superconducting order parameter self-consistently, where  $F(\Delta,q)$ represents the free energy density of the Floquet Hamiltonian $\mathcal{H}_F$, given by 
\begin{equation}
F(q,\Delta) = -\frac{1}{L\beta}\sum_{m} \ln\left[1+e^{-\beta E_{m}(q)}\right] + \frac{|\Delta(q)|^2}{U}\ .
\label{eqn:free_energy}
\end{equation}
where, $\beta^{-1} = k_B T$ denotes the temperature, and $m$ labels the Floquet quasi-energy bands. Using the self-consistent solution $\Delta(q)$, we can compute the net supercurrent density $I(q)$, following Eq.~(\ref{eqn:current}), from the knowledge of self-consistent Floquet-BdG eigenstate. 

The quasienergy spectrum, obtained employing the fully self-consistent solution $\Delta(q)$, is depicted as a function of $q$ in Figs.~\ref{fig:Fig2}(a-c) 
for three representative parameter sets revealing distinct realizations of Floquet TSC signatures. In Fig.~\ref{fig:Fig2}(a), the spectrum solely exhibits topological $\pi$-FMMs exclusively in the negative-$q$ sector, while Fig.~\ref{fig:Fig2}(b) supports the emergence of only $0$-FMMs in the positive-$q$ sector. Notably, in Fig.~\ref{fig:Fig2}(c), both sectors are realized simultaneously, with $0$-FMMs in the positive-$q$ region and $\pi$-FMMs in the negative-$q$ region, evidently highlighting the switching between topological $0$- and $\pi$-FMMs upon sign-reversal of $q$. The topological regimes are further characterized with $P_x=0.5$ for the respective spectrum as a function of $q$ as shown in Figs.~\ref{fig:Fig2}(a-c). These observations, obtained from self-consistent analysis, 
are consistent with the spectral analysis discussed earlier within the non-self-consistent framework (see Fig.~\ref{fig:Fig1}(c)), indicating the qualitative viability of the non-self-consistent approach. 

{\textcolor{blue}{\textit{Interplay between supercurrent and topological FMMs-}}} The corresponding supercurrent density $I(q)$ is shown in Figs.~\ref{fig:Fig2}(e-g), illustrating its functional dependence on $q$ at different temperatures $\beta^{-1}$. A clear correspondence emerges between the topological regime, hosting FMMs in the quasi-spectrum, and the current response within that above window. Interestingly, the $0$- and $\pi$-FMMs result in opposite flow of supercurrents which is also consistent with positive and negative values of $q$. Such a correspondence clearly demonstrates a supercurrent-induced topological transition and tunable switching between topological $0$- and $\pi$-FMMs via controlling the supercurrent direction.
Furthermore, $I(q)$ profiles reveal distinct signatures across the topological regimes for different values of $q$. At zero temperature, the current exhibits an abrupt change upon entering or exiting the topological phase (especially for $0$-FMMs). To be precise, there is a change in slopes in $I(q)$ across the topological boundaries with respect to $q$. This behavior is directly linked to the nature of the self-consistent solution of the superconducting order parameter profile $\Delta(q)$ as depicted in Fig.~\ref{fig:Fig2}(d).  

Note that, in Fig.~\ref{fig:Fig2}(d), $\Delta(q)$ profile embeds $0$- and $\pi$-FMM simultaneously as shown in Figs.~\ref{fig:Fig2}(c,g). The zero-temperature profile of $\Delta(q)$ exhibits a sharp suppression in the topological regime $0.28<q<0.5$, hosting $0$-FMMs (see the blue curve in Fig.~\ref{fig:Fig2}(d)), 
which is accompanied by a reduced magnitude of supercurrent. Interestingly, current magnitude is higher for the regime $-0.3<q<-0.15$, supporting $\pi$-FMMs, within which the order parameter remains nearly constant. At finite temperatures, the suppression of $\Delta$ near the topological phase transition becomes progressively smoother (see the green curve in Fig.~\ref{fig:Fig2}(d)). Consequently, the current profile demonstrates a gradual variation as can be observed 
in Figs.~\ref{fig:Fig2}(f,g). Importantly, the current retains an approximately linear dependence on $q$ throughout the entire topological regimes, with an evident change in slope across the topological boundary (see Figs.~\ref{fig:Fig2}(e-g)). The corresponding change in the magnitude and sign of slope in terms 
of the current profile provides a direct signature of the topological $0$- and $\pi$- FMMs. This profile qualitatively remains similar as the temperature increases, however, becomes smoother due to thermal fluctuations as depicted in Figs.~\ref{fig:Fig2}(e-g).

\begin{figure}[t]
\centering
\includegraphics[width=1\linewidth]{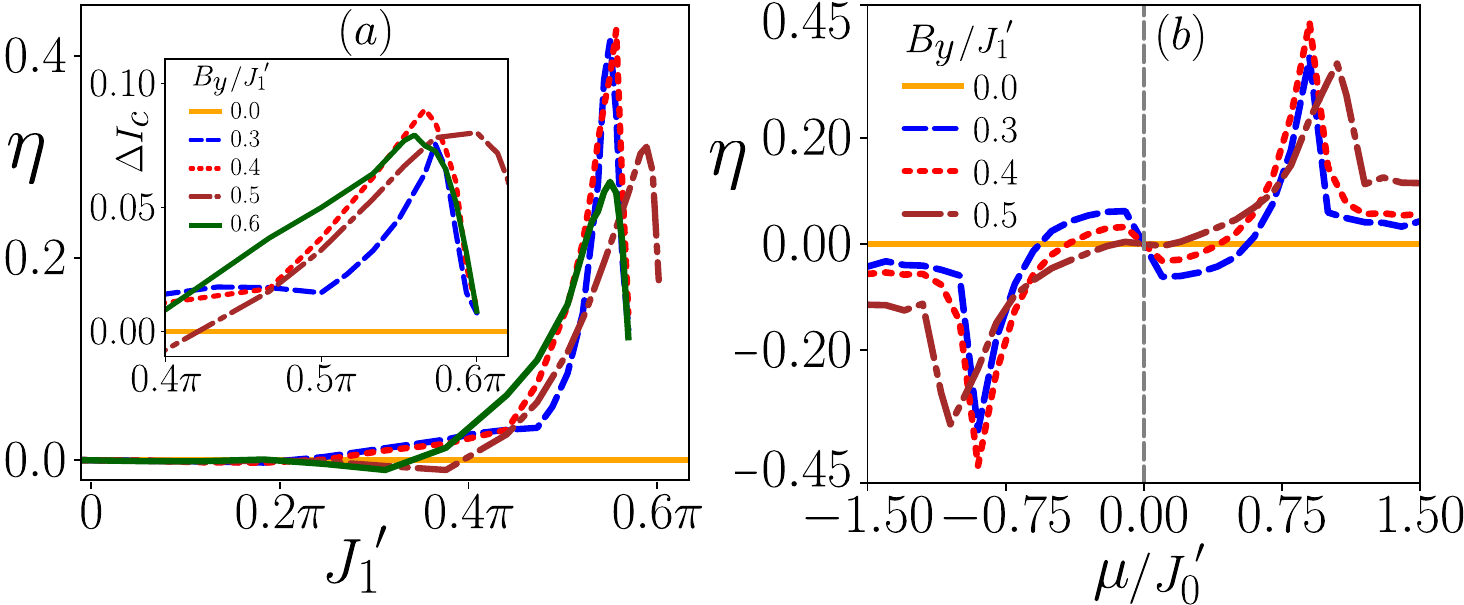}
\caption{
We showcase the profile of diode efficiency $\eta$, considering different values of $B_y$, as a function of ${J'_1}$ for $\mu/{J'_0}=1.2$ in panel (a). 
The corresponding critical current asymmetry $\Delta I_c$ is shown in the inset of panel (a). On the other hand, $\eta$ is depicted as a function of $\mu/{J_0}$ choosing an optimal value ${J'_1}=0.55 \pi$. The other system parameters are given as $(t,\alpha)=1 {J'_0},\beta^{-1}/\Delta_0=0.1, {J'_0} = 0.45 \pi$.}
\label{fig:Fig3}
\end{figure}	

{\textcolor{blue}{\textit{Non-reciprocal charge transport (SDE)-}}} Here, we focus on the nonreciprocal transport characteristics encoded in the current response $I(q)$ as  depicted Figs.~\ref{fig:Fig2}(e)-(g). The current profiles are intrinsically asymmetric with respect to $q=0$, satisfying $I(q)\neq I(-q)$ in the presence of finite $B_y$. Such nonreciprocal transport properties in semiconductor-superconductor hybrid systems has been widely associated with the FF pairing mechanism, originated from a finite Cooper pair momentum $q_0$ in superconducting ground state or equilibrium state in absence of external supercurrent i.e.,  $I{(q_0)}=0$~\cite{LiangPNAS,SLlicprl,Yanasediode,Picoli,Nagaosanjp, bhowmik_2025}. The asymmetric current profiles are shown in Figs.~\ref{fig:Fig2}(e)-(g). 
It is evident that a finite shift of the Cooper pair momentum by an amount $q_0$ for which $I(q_0)=0$ (see the inset of Fig.~\ref{fig:Fig2}(f)). We find that 
the finite momentum $q_0$ of the FF state increases with driving parameter $J'_1$, irrespectice of the choice of $B_y \ne 0$ as depicted in  Fig.~\ref{fig:Fig2}(h). Note that, a finite $B_y$ is required to achieve a finite $q_0$, establishing the crucial role of $B_y$ in stabilizing the Floquet drive-induced FF superconducting phase. The corresponding results for the $B_y=0$ case, exhibiting reciprocal quasi-energy spectra and supercurrent response, 
are discussed in the Supplemental Material~\cite{supp}. 

The non-reciprocity is further accompanied by the asymmetry of critical supercurrents in opposite flow directions: $I^{+}_c\ne-I^{-}_c$ and quantified by the diode efficiency $\eta={(|I_{c}^+|-|I_{c}^-|)}/{(|I_{c}^+|+|I_{c}^-|)}$~\cite{Lossdiode,LiangPNAS,Qu2013natcom}. Here, $I_c^{\pm}$ denote the maximum characteristic current of a superconductor, exceeding which leads to the de-pairing of superconductivity to normal Ohmic state. The behavior of $\eta$ is depicted as a function of the Floquet driving strength ${J'_1}$ and $\mu$ in Figs.~\ref{fig:Fig3}(a,b), respectively, for different choices of $B_y$. Additionally, we also demonstrate the variation of the critical supercurrent asymmetry $\Delta I_c= |I_{c}^+|-|I_{c}^-|$, quantifying the measure of non-reciprocity, as a function of ${J'_1}$ (see the inset of Fig.~\ref{fig:Fig3}(a)). Given the fact that the diode efficiency $\eta$ serves as a standard measure of SDE performance, the quantity $\Delta I_c$ is particularly relevant from an experimental and application-oriented perspective, where enhanced asymmetry between opposite critical currents is particularly relevant~\cite{Jaime_2023}. Finally, the optimal value of ${J'_1}$, corresponding to the maximum critical current asymmetry $\Delta I_c$, is further employed to evaluate the diode efficiency as a function of the chemical potential $\mu$ as shown in Fig.~\ref{fig:Fig3}(b). This analysis allows us to estimate the performance of the Floquet-engineered SDE in terms of the driving parameters. This leads to a highest diode efficiency of $\eta\approx45\%$ within the present setup.       

{\textcolor{blue}{\textit{Summary and Conclusions-}}} In the static model of Rashba nanowire proximitized to a $s$-wave superconductor, it has been
shown that in-plane magnetic field leads to Fermi-surface anisotropy, allowing for a possibe finite-momentum pairing of electrons, while perpendicular magnetic field causes topological modes to emerge. Using these concepts, we consider the above model in the presence of external supercurrent to examine the emergence of FF superconducting state in the quasi-energy spectrum under the application of periodic drive while the static limit does not support any FF ground state. 
This setup enables us to explore the interplay between Floquet TSC, finite-momentum pairing, and nonreciprocal transport leading to SDE. We find that the Floquet spectral properties are intertwined with supercurrent leading to $0$- and $\pi$- FMMs for postive and negative values of supercurrent. This is an example of  supercurrent-driven topological transition where the non-reciprocity of the supercurrent is manisfested leading to SDE. We validate the above findings following a self-consistent maen-field calculations confirming the appearance of Floquet-induced FF superconducting state with pronounced nonreciprocal supercurrent responses and topological signatures. One can parametrically tune the driving amplitudes in terms of the magnetic field to maximize the efficiency as a function of chemical potential. Given experimental advancement on Floquet techniques in solid-state systems~\cite{wang2013observation,Mahmood2016,McIver2020,Basov2017} 
and SDE~\cite{ando2020observation,lin2022zero}, our theoretical proposal may be possible to realize 
in heterostructures involving common $s$-wave superconductors and strong 
SOC materials~\cite{Mourik}.


{\textcolor{blue}{\textit{Data Availibility Statement}}} \--- The datasets generated and analyzed during the current study are available from the authors upon reasonable request.

{\textcolor{blue}{\textit{Acknowledgments}}} \---  S.B. and A.S. acknowledge the SAMKHYA: HPC Facility provided at IOP, Bhubaneswar and the two workstations provided by IOP, Bhubaneshwar from  DAE APEX project for the numerical computations.  T.N. and A.S. thank the Advanced Research Grant (ARG) from Anusandhan National Research Foundation Grant No. ANRF/ARG/2025/003163/PS.

\bibliography{ref.bib}		

\begin{thebibliography}{61}%
\makeatletter
\providecommand \@ifxundefined [1]{%
 \@ifx{#1\undefined}
}%
\providecommand \@ifnum [1]{%
 \ifnum #1\expandafter \@firstoftwo
 \else \expandafter \@secondoftwo
 \fi
}%
\providecommand \@ifx [1]{%
 \ifx #1\expandafter \@firstoftwo
 \else \expandafter \@secondoftwo
 \fi
}%
\providecommand \natexlab [1]{#1}%
\providecommand \enquote  [1]{``#1''}%
\providecommand \bibnamefont  [1]{#1}%
\providecommand \bibfnamefont [1]{#1}%
\providecommand \citenamefont [1]{#1}%
\providecommand \href@noop [0]{\@secondoftwo}%
\providecommand \href [0]{\begingroup \@sanitize@url \@href}%
\providecommand \@href[1]{\@@startlink{#1}\@@href}%
\providecommand \@@href[1]{\endgroup#1\@@endlink}%
\providecommand \@sanitize@url [0]{\catcode `\\12\catcode `\$12\catcode
  `\&12\catcode `\#12\catcode `\^12\catcode `\_12\catcode `\%12\relax}%
\providecommand \@@startlink[1]{}%
\providecommand \@@endlink[0]{}%
\providecommand \url  [0]{\begingroup\@sanitize@url \@url }%
\providecommand \@url [1]{\endgroup\@href {#1}{\urlprefix }}%
\providecommand \urlprefix  [0]{URL }%
\providecommand \Eprint [0]{\href }%
\providecommand \doibase [0]{http://dx.doi.org/}%
\providecommand \selectlanguage [0]{\@gobble}%
\providecommand \bibinfo  [0]{\@secondoftwo}%
\providecommand \bibfield  [0]{\@secondoftwo}%
\providecommand \translation [1]{[#1]}%
\providecommand \BibitemOpen [0]{}%
\providecommand \bibitemStop [0]{}%
\providecommand \bibitemNoStop [0]{.\EOS\space}%
\providecommand \EOS [0]{\spacefactor3000\relax}%
\providecommand \BibitemShut  [1]{\csname bibitem#1\endcsname}%
\let\auto@bib@innerbib\@empty
\bibitem [{\citenamefont {Ivanov}(2001)}]{Ivanov}%
  \BibitemOpen
  \bibfield  {author} {\bibinfo {author} {\bibfnamefont {D.~A.}\ \bibnamefont
  {Ivanov}},\ }\bibfield  {title} {\enquote {\bibinfo {title} {Non-abelian
  statistics of half-quantum vortices in $\mathit{p}$-wave superconductors},}\
  }\href {\doibase 10.1103/PhysRevLett.86.268} {\bibfield  {journal} {\bibinfo
  {journal} {Phys. Rev. Lett.}\ }\textbf {\bibinfo {volume} {86}},\ \bibinfo
  {pages} {268--271} (\bibinfo {year} {2001})}\BibitemShut {NoStop}%
\bibitem [{\citenamefont {Kitaev}(2003)}]{KITAEV20032}%
  \BibitemOpen
  \bibfield  {author} {\bibinfo {author} {\bibfnamefont {A.Yu.}\ \bibnamefont
  {Kitaev}},\ }\bibfield  {title} {\enquote {\bibinfo {title} {Fault-tolerant
  quantum computation by anyons},}\ }\href {\doibase
  https://doi.org/10.1016/S0003-4916(02)00018-0} {\bibfield  {journal}
  {\bibinfo  {journal} {Annals of Physics}\ }\textbf {\bibinfo {volume}
  {303}},\ \bibinfo {pages} {2--30} (\bibinfo {year} {2003})}\BibitemShut
  {NoStop}%
\bibitem [{\citenamefont {Stern}(2010)}]{Stern2010}%
  \BibitemOpen
  \bibfield  {author} {\bibinfo {author} {\bibfnamefont {Ady}\ \bibnamefont
  {Stern}},\ }\bibfield  {title} {\enquote {\bibinfo {title} {Non-abelian
  states of matter},}\ }\href {\doibase 10.1038/nature08915} {\bibfield
  {journal} {\bibinfo  {journal} {Nature}\ }\textbf {\bibinfo {volume} {464}},\
  \bibinfo {pages} {187--193} (\bibinfo {year} {2010})}\BibitemShut {NoStop}%
\bibitem [{\citenamefont {Nayak}\ \emph {et~al.}(2008)\citenamefont {Nayak},
  \citenamefont {Simon}, \citenamefont {Stern}, \citenamefont {Freedman},\ and\
  \citenamefont {Das~Sarma}}]{CNayak}%
  \BibitemOpen
  \bibfield  {author} {\bibinfo {author} {\bibfnamefont {Chetan}\ \bibnamefont
  {Nayak}}, \bibinfo {author} {\bibfnamefont {Steven~H.}\ \bibnamefont
  {Simon}}, \bibinfo {author} {\bibfnamefont {Ady}\ \bibnamefont {Stern}},
  \bibinfo {author} {\bibfnamefont {Michael}\ \bibnamefont {Freedman}}, \ and\
  \bibinfo {author} {\bibfnamefont {Sankar}\ \bibnamefont {Das~Sarma}},\
  }\bibfield  {title} {\enquote {\bibinfo {title} {Non-abelian anyons and
  topological quantum computation},}\ }\href {\doibase
  10.1103/RevModPhys.80.1083} {\bibfield  {journal} {\bibinfo  {journal} {Rev.
  Mod. Phys.}\ }\textbf {\bibinfo {volume} {80}},\ \bibinfo {pages}
  {1083--1159} (\bibinfo {year} {2008})}\BibitemShut {NoStop}%
\bibitem [{\citenamefont {Kitaev}(2009)}]{Kitaev2009}%
  \BibitemOpen
  \bibfield  {author} {\bibinfo {author} {\bibfnamefont {Alexei}\ \bibnamefont
  {Kitaev}},\ }\bibfield  {title} {\enquote {\bibinfo {title} {Periodic table
  for topological insulators and superconductors},}\ }\href {\doibase
  10.1063/1.3149495} {\bibfield  {journal} {\bibinfo  {journal} {AIP Conference
  Proceedings}\ }\textbf {\bibinfo {volume} {1134}},\ \bibinfo {pages} {22--30}
  (\bibinfo {year} {2009})}\BibitemShut {NoStop}%
\bibitem [{\citenamefont {Kitaev}(2001)}]{Kitaev_2001}%
  \BibitemOpen
  \bibfield  {author} {\bibinfo {author} {\bibfnamefont {A~Yu}\ \bibnamefont
  {Kitaev}},\ }\bibfield  {title} {\enquote {\bibinfo {title} {Unpaired
  majorana fermions in quantumwires},}\ }\href {\doibase
  10.1070/1063-7869/44/10S/S29} {\bibfield  {journal} {\bibinfo  {journal}
  {Physics-Uspekhi}\ }\textbf {\bibinfo {volume} {44}},\ \bibinfo {pages} {131}
  (\bibinfo {year} {2001})}\BibitemShut {NoStop}%
\bibitem [{\citenamefont {Leijnse}\ and\ \citenamefont
  {Flensberg}(2012)}]{Leijnse_2012}%
  \BibitemOpen
  \bibfield  {author} {\bibinfo {author} {\bibfnamefont {Martin}\ \bibnamefont
  {Leijnse}}\ and\ \bibinfo {author} {\bibfnamefont {Karsten}\ \bibnamefont
  {Flensberg}},\ }\bibfield  {title} {\enquote {\bibinfo {title} {Introduction
  to topological superconductivity and majorana fermions},}\ }\href {\doibase
  10.1088/0268-1242/27/12/124003} {\bibfield  {journal} {\bibinfo  {journal}
  {Semiconductor Science and Technology}\ }\textbf {\bibinfo {volume} {27}},\
  \bibinfo {pages} {124003} (\bibinfo {year} {2012})}\BibitemShut {NoStop}%
\bibitem [{\citenamefont {Aguado}(2017)}]{Aguado2017}%
  \BibitemOpen
  \bibfield  {author} {\bibinfo {author} {\bibfnamefont {Ramón}\ \bibnamefont
  {Aguado}},\ }\bibfield  {title} {\enquote {\bibinfo {title} {Majorana
  quasiparticles in condensed matter},}\ }\href {\doibase
  10.1393/ncr/i2017-10141-9} {\bibfield  {journal} {\bibinfo  {journal} {La
  Rivista del Nuovo Cimento}\ }\textbf {\bibinfo {volume} {40}},\ \bibinfo
  {pages} {523--593} (\bibinfo {year} {2017})}\BibitemShut {NoStop}%
\bibitem [{\citenamefont {Alicea}(2012)}]{Alicea_2012}%
  \BibitemOpen
  \bibfield  {author} {\bibinfo {author} {\bibfnamefont {Jason}\ \bibnamefont
  {Alicea}},\ }\bibfield  {title} {\enquote {\bibinfo {title} {New directions
  in the pursuit of majorana fermions in solid state systems},}\ }\href
  {\doibase 10.1088/0034-4885/75/7/076501} {\bibfield  {journal} {\bibinfo
  {journal} {Reports on Progress in Physics}\ }\textbf {\bibinfo {volume}
  {75}},\ \bibinfo {pages} {076501} (\bibinfo {year} {2012})}\BibitemShut
  {NoStop}%
\bibitem [{\citenamefont {Lutchyn}\ \emph {et~al.}(2010)\citenamefont
  {Lutchyn}, \citenamefont {Sau},\ and\ \citenamefont
  {Das~Sarma}}]{Lutchyn_Sau}%
  \BibitemOpen
  \bibfield  {author} {\bibinfo {author} {\bibfnamefont {Roman~M.}\
  \bibnamefont {Lutchyn}}, \bibinfo {author} {\bibfnamefont {Jay~D.}\
  \bibnamefont {Sau}}, \ and\ \bibinfo {author} {\bibfnamefont
  {S.}~\bibnamefont {Das~Sarma}},\ }\bibfield  {title} {\enquote {\bibinfo
  {title} {Majorana fermions and a topological phase transition in
  semiconductor-superconductor heterostructures},}\ }\href {\doibase
  10.1103/PhysRevLett.105.077001} {\bibfield  {journal} {\bibinfo  {journal}
  {Phys. Rev. Lett.}\ }\textbf {\bibinfo {volume} {105}},\ \bibinfo {pages}
  {077001} (\bibinfo {year} {2010})}\BibitemShut {NoStop}%
\bibitem [{\citenamefont {Mourik}\ \emph {et~al.}(2012)\citenamefont {Mourik},
  \citenamefont {Zuo}, \citenamefont {Frolov}, \citenamefont {Plissard},
  \citenamefont {Bakkers},\ and\ \citenamefont {Kouwenhoven}}]{Mourik}%
  \BibitemOpen
  \bibfield  {author} {\bibinfo {author} {\bibfnamefont {V.}~\bibnamefont
  {Mourik}}, \bibinfo {author} {\bibfnamefont {K.}~\bibnamefont {Zuo}},
  \bibinfo {author} {\bibfnamefont {S.~M.}\ \bibnamefont {Frolov}}, \bibinfo
  {author} {\bibfnamefont {S.~R.}\ \bibnamefont {Plissard}}, \bibinfo {author}
  {\bibfnamefont {E.~P. A.~M.}\ \bibnamefont {Bakkers}}, \ and\ \bibinfo
  {author} {\bibfnamefont {L.~P.}\ \bibnamefont {Kouwenhoven}},\ }\bibfield
  {title} {\enquote {\bibinfo {title} {Signatures of majorana fermions in
  hybrid superconductor-semiconductor nanowire devices},}\ }\href {\doibase
  10.1126/science.1222360} {\bibfield  {journal} {\bibinfo  {journal}
  {Science}\ }\textbf {\bibinfo {volume} {336}},\ \bibinfo {pages} {1003--1007}
  (\bibinfo {year} {2012})}\BibitemShut {NoStop}%
\bibitem [{\citenamefont {Qi}\ and\ \citenamefont {Zhang}(2011)}]{Zhang_Qi}%
  \BibitemOpen
  \bibfield  {author} {\bibinfo {author} {\bibfnamefont {Xiao-Liang}\
  \bibnamefont {Qi}}\ and\ \bibinfo {author} {\bibfnamefont {Shou-Cheng}\
  \bibnamefont {Zhang}},\ }\bibfield  {title} {\enquote {\bibinfo {title}
  {Topological insulators and superconductors},}\ }\href {\doibase
  10.1103/RevModPhys.83.1057} {\bibfield  {journal} {\bibinfo  {journal} {Rev.
  Mod. Phys.}\ }\textbf {\bibinfo {volume} {83}},\ \bibinfo {pages}
  {1057--1110} (\bibinfo {year} {2011})}\BibitemShut {NoStop}%
\bibitem [{\citenamefont {Oreg}\ \emph {et~al.}(2010)\citenamefont {Oreg},
  \citenamefont {Refael},\ and\ \citenamefont {von Oppen}}]{Yuval_Oreg_Oppen}%
  \BibitemOpen
  \bibfield  {author} {\bibinfo {author} {\bibfnamefont {Yuval}\ \bibnamefont
  {Oreg}}, \bibinfo {author} {\bibfnamefont {Gil}\ \bibnamefont {Refael}}, \
  and\ \bibinfo {author} {\bibfnamefont {Felix}\ \bibnamefont {von Oppen}},\
  }\bibfield  {title} {\enquote {\bibinfo {title} {Helical liquids and majorana
  bound states in quantum wires},}\ }\href {\doibase
  10.1103/PhysRevLett.105.177002} {\bibfield  {journal} {\bibinfo  {journal}
  {Phys. Rev. Lett.}\ }\textbf {\bibinfo {volume} {105}},\ \bibinfo {pages}
  {177002} (\bibinfo {year} {2010})}\BibitemShut {NoStop}%
\bibitem [{\citenamefont {Beenakker}(2013)}]{Beenakker}%
  \BibitemOpen
  \bibfield  {author} {\bibinfo {author} {\bibfnamefont {C.W.J.}\ \bibnamefont
  {Beenakker}},\ }\bibfield  {title} {\enquote {\bibinfo {title} {Majorana
  fermions in superconductors},}\ }\href {\doibase
  10.1146/annurev-conmatphys-030212-184337} {\bibfield  {journal} {\bibinfo
  {journal} {Annual Review of Condensed Matter Physics}\ }\textbf {\bibinfo
  {volume} {4}},\ \bibinfo {pages} {113--136} (\bibinfo {year}
  {2013})}\BibitemShut {NoStop}%
\bibitem [{\citenamefont {Sau}\ \emph {et~al.}(2010)\citenamefont {Sau},
  \citenamefont {Lutchyn}, \citenamefont {Tewari},\ and\ \citenamefont
  {Das~Sarma}}]{SAU_PRL}%
  \BibitemOpen
  \bibfield  {author} {\bibinfo {author} {\bibfnamefont {Jay~D.}\ \bibnamefont
  {Sau}}, \bibinfo {author} {\bibfnamefont {Roman~M.}\ \bibnamefont {Lutchyn}},
  \bibinfo {author} {\bibfnamefont {Sumanta}\ \bibnamefont {Tewari}}, \ and\
  \bibinfo {author} {\bibfnamefont {S.}~\bibnamefont {Das~Sarma}},\ }\bibfield
  {title} {\enquote {\bibinfo {title} {Generic new platform for topological
  quantum computation using semiconductor heterostructures},}\ }\href {\doibase
  10.1103/PhysRevLett.104.040502} {\bibfield  {journal} {\bibinfo  {journal}
  {Phys. Rev. Lett.}\ }\textbf {\bibinfo {volume} {104}},\ \bibinfo {pages}
  {040502} (\bibinfo {year} {2010})}\BibitemShut {NoStop}%
\bibitem [{\citenamefont {Tewari}\ and\ \citenamefont {Sau}(2012)}]{PRL_SAU}%
  \BibitemOpen
  \bibfield  {author} {\bibinfo {author} {\bibfnamefont {Sumanta}\ \bibnamefont
  {Tewari}}\ and\ \bibinfo {author} {\bibfnamefont {Jay~D.}\ \bibnamefont
  {Sau}},\ }\bibfield  {title} {\enquote {\bibinfo {title} {Topological
  invariants for spin-orbit coupled superconductor nanowires},}\ }\href
  {\doibase 10.1103/PhysRevLett.109.150408} {\bibfield  {journal} {\bibinfo
  {journal} {Phys. Rev. Lett.}\ }\textbf {\bibinfo {volume} {109}},\ \bibinfo
  {pages} {150408} (\bibinfo {year} {2012})}\BibitemShut {NoStop}%
\bibitem [{\citenamefont {Rainis}\ \emph {et~al.}(2013)\citenamefont {Rainis},
  \citenamefont {Trifunovic}, \citenamefont {Klinovaja},\ and\ \citenamefont
  {Loss}}]{Rainis}%
  \BibitemOpen
  \bibfield  {author} {\bibinfo {author} {\bibfnamefont {Diego}\ \bibnamefont
  {Rainis}}, \bibinfo {author} {\bibfnamefont {Luka}\ \bibnamefont
  {Trifunovic}}, \bibinfo {author} {\bibfnamefont {Jelena}\ \bibnamefont
  {Klinovaja}}, \ and\ \bibinfo {author} {\bibfnamefont {Daniel}\ \bibnamefont
  {Loss}},\ }\bibfield  {title} {\enquote {\bibinfo {title} {Towards a
  realistic transport modeling in a superconducting nanowire with majorana
  fermions},}\ }\href {\doibase 10.1103/PhysRevB.87.024515} {\bibfield
  {journal} {\bibinfo  {journal} {Phys. Rev. B}\ }\textbf {\bibinfo {volume}
  {87}},\ \bibinfo {pages} {024515} (\bibinfo {year} {2013})}\BibitemShut
  {NoStop}%
\bibitem [{\citenamefont {Oka}\ and\ \citenamefont
  {Aoki}(2009{\natexlab{a}})}]{OKA2009}%
  \BibitemOpen
  \bibfield  {author} {\bibinfo {author} {\bibfnamefont {Takashi}\ \bibnamefont
  {Oka}}\ and\ \bibinfo {author} {\bibfnamefont {Hideo}\ \bibnamefont {Aoki}},\
  }\bibfield  {title} {\enquote {\bibinfo {title} {Photovoltaic hall effect in
  graphene},}\ }\href {\doibase 10.1103/PhysRevB.79.081406} {\bibfield
  {journal} {\bibinfo  {journal} {Phys. Rev. B}\ }\textbf {\bibinfo {volume}
  {79}},\ \bibinfo {pages} {081406(R)} (\bibinfo {year}
  {2009}{\natexlab{a}})}\BibitemShut {NoStop}%
\bibitem [{\citenamefont {Lindner}\ \emph {et~al.}(2011)\citenamefont
  {Lindner}, \citenamefont {Refael},\ and\ \citenamefont
  {Galitski}}]{lindner2011floquet}%
  \BibitemOpen
  \bibfield  {author} {\bibinfo {author} {\bibfnamefont {Netanel~H}\
  \bibnamefont {Lindner}}, \bibinfo {author} {\bibfnamefont {Gil}\ \bibnamefont
  {Refael}}, \ and\ \bibinfo {author} {\bibfnamefont {Victor}\ \bibnamefont
  {Galitski}},\ }\bibfield  {title} {\enquote {\bibinfo {title} {Floquet
  topological insulator in semiconductor quantum wells},}\ }\href {\doibase
  10.1038/nphys1926} {\bibfield  {journal} {\bibinfo  {journal} {Nature
  Physics}\ }\textbf {\bibinfo {volume} {7}},\ \bibinfo {pages} {490--495}
  (\bibinfo {year} {2011})}\BibitemShut {NoStop}%
\bibitem [{\citenamefont {Farrell}\ and\ \citenamefont
  {Pereg-Barnea}(2015)}]{Barnea}%
  \BibitemOpen
  \bibfield  {author} {\bibinfo {author} {\bibfnamefont {Aaron}\ \bibnamefont
  {Farrell}}\ and\ \bibinfo {author} {\bibfnamefont {T.}~\bibnamefont
  {Pereg-Barnea}},\ }\bibfield  {title} {\enquote {\bibinfo {title}
  {Photon-inhibited topological transport in quantum well heterostructures},}\
  }\href {\doibase 10.1103/PhysRevLett.115.106403} {\bibfield  {journal}
  {\bibinfo  {journal} {Phys. Rev. Lett.}\ }\textbf {\bibinfo {volume} {115}},\
  \bibinfo {pages} {106403} (\bibinfo {year} {2015})}\BibitemShut {NoStop}%
\bibitem [{\citenamefont {Mondal}\ \emph
  {et~al.}(2023{\natexlab{a}})\citenamefont {Mondal}, \citenamefont {Ghosh},
  \citenamefont {Nag},\ and\ \citenamefont {Saha}}]{Mondal_1}%
  \BibitemOpen
  \bibfield  {author} {\bibinfo {author} {\bibfnamefont {Debashish}\
  \bibnamefont {Mondal}}, \bibinfo {author} {\bibfnamefont {Arnob~Kumar}\
  \bibnamefont {Ghosh}}, \bibinfo {author} {\bibfnamefont {Tanay}\ \bibnamefont
  {Nag}}, \ and\ \bibinfo {author} {\bibfnamefont {Arijit}\ \bibnamefont
  {Saha}},\ }\bibfield  {title} {\enquote {\bibinfo {title} {Engineering
  anomalous floquet majorana modes and their time evolution in a helical shiba
  chain},}\ }\href {\doibase 10.1103/PhysRevB.108.L081403} {\bibfield
  {journal} {\bibinfo  {journal} {Phys. Rev. B}\ }\textbf {\bibinfo {volume}
  {108}},\ \bibinfo {pages} {L081403} (\bibinfo {year}
  {2023}{\natexlab{a}})}\BibitemShut {NoStop}%
\bibitem [{\citenamefont {Mondal}\ \emph
  {et~al.}(2023{\natexlab{b}})\citenamefont {Mondal}, \citenamefont {Ghosh},
  \citenamefont {Nag},\ and\ \citenamefont {Saha}}]{Mondal2}%
  \BibitemOpen
  \bibfield  {author} {\bibinfo {author} {\bibfnamefont {Debashish}\
  \bibnamefont {Mondal}}, \bibinfo {author} {\bibfnamefont {Arnob~Kumar}\
  \bibnamefont {Ghosh}}, \bibinfo {author} {\bibfnamefont {Tanay}\ \bibnamefont
  {Nag}}, \ and\ \bibinfo {author} {\bibfnamefont {Arijit}\ \bibnamefont
  {Saha}},\ }\bibfield  {title} {\enquote {\bibinfo {title} {Topological
  characterization and stability of floquet majorana modes in rashba
  nanowires},}\ }\href {\doibase 10.1103/PhysRevB.107.035427} {\bibfield
  {journal} {\bibinfo  {journal} {Phys. Rev. B}\ }\textbf {\bibinfo {volume}
  {107}},\ \bibinfo {pages} {035427} (\bibinfo {year}
  {2023}{\natexlab{b}})}\BibitemShut {NoStop}%
\bibitem [{\citenamefont {Eckardt}(2017)}]{Eckardt}%
  \BibitemOpen
  \bibfield  {author} {\bibinfo {author} {\bibfnamefont {Andr\'e}\ \bibnamefont
  {Eckardt}},\ }\bibfield  {title} {\enquote {\bibinfo {title} {Colloquium:
  Atomic quantum gases in periodically driven optical lattices},}\ }\href
  {\doibase 10.1103/RevModPhys.89.011004} {\bibfield  {journal} {\bibinfo
  {journal} {Rev. Mod. Phys.}\ }\textbf {\bibinfo {volume} {89}},\ \bibinfo
  {pages} {011004} (\bibinfo {year} {2017})}\BibitemShut {NoStop}%
\bibitem [{\citenamefont {Ghosh}\ \emph {et~al.}(2024)\citenamefont {Ghosh},
  \citenamefont {Nag},\ and\ \citenamefont {Saha}}]{ghosh2024generation}%
  \BibitemOpen
  \bibfield  {author} {\bibinfo {author} {\bibfnamefont {Arnob~Kumar}\
  \bibnamefont {Ghosh}}, \bibinfo {author} {\bibfnamefont {Tanay}\ \bibnamefont
  {Nag}}, \ and\ \bibinfo {author} {\bibfnamefont {Arijit}\ \bibnamefont
  {Saha}},\ }\bibfield  {title} {\enquote {\bibinfo {title} {Generation of
  higher-order topological insulators using periodic driving},}\ }\href
  {\doibase 10.1088/1361-648X/ad0e2d} {\bibfield  {journal} {\bibinfo
  {journal} {Journal of Physics: Condensed Matter}\ }\textbf {\bibinfo {volume}
  {36}},\ \bibinfo {pages} {093001} (\bibinfo {year} {2024})}\BibitemShut
  {NoStop}%
\bibitem [{\citenamefont {Ghosh}\ \emph {et~al.}(2022)\citenamefont {Ghosh},
  \citenamefont {Nag},\ and\ \citenamefont {Saha}}]{Ghosh2022}%
  \BibitemOpen
  \bibfield  {author} {\bibinfo {author} {\bibfnamefont {Arnob~Kumar}\
  \bibnamefont {Ghosh}}, \bibinfo {author} {\bibfnamefont {Tanay}\ \bibnamefont
  {Nag}}, \ and\ \bibinfo {author} {\bibfnamefont {Arijit}\ \bibnamefont
  {Saha}},\ }\bibfield  {title} {\enquote {\bibinfo {title} {Dynamical
  construction of quadrupolar and octupolar topological superconductors},}\
  }\href {\doibase 10.1103/PhysRevB.105.155406} {\bibfield  {journal} {\bibinfo
   {journal} {Phys. Rev. B}\ }\textbf {\bibinfo {volume} {105}},\ \bibinfo
  {pages} {155406} (\bibinfo {year} {2022})}\BibitemShut {NoStop}%
\bibitem [{\citenamefont {Rudner}\ \emph {et~al.}(2013)\citenamefont {Rudner},
  \citenamefont {Lindner}, \citenamefont {Berg},\ and\ \citenamefont
  {Levin}}]{Rudner2013}%
  \BibitemOpen
  \bibfield  {author} {\bibinfo {author} {\bibfnamefont {Mark~S.}\ \bibnamefont
  {Rudner}}, \bibinfo {author} {\bibfnamefont {Netanel~H.}\ \bibnamefont
  {Lindner}}, \bibinfo {author} {\bibfnamefont {Erez}\ \bibnamefont {Berg}}, \
  and\ \bibinfo {author} {\bibfnamefont {Michael}\ \bibnamefont {Levin}},\
  }\bibfield  {title} {\enquote {\bibinfo {title} {Anomalous edge states and
  the bulk-edge correspondence for periodically driven two-dimensional
  systems},}\ }\href {\doibase 10.1103/PhysRevX.3.031005} {\bibfield  {journal}
  {\bibinfo  {journal} {Phys. Rev. X}\ }\textbf {\bibinfo {volume} {3}},\
  \bibinfo {pages} {031005} (\bibinfo {year} {2013})}\BibitemShut {NoStop}%
\bibitem [{\citenamefont {Yao}\ \emph {et~al.}(2017)\citenamefont {Yao},
  \citenamefont {Yan},\ and\ \citenamefont {Wang}}]{Yao2017}%
  \BibitemOpen
  \bibfield  {author} {\bibinfo {author} {\bibfnamefont {Shunyu}\ \bibnamefont
  {Yao}}, \bibinfo {author} {\bibfnamefont {Zhongbo}\ \bibnamefont {Yan}}, \
  and\ \bibinfo {author} {\bibfnamefont {Zhong}\ \bibnamefont {Wang}},\
  }\bibfield  {title} {\enquote {\bibinfo {title} {Topological invariants of
  floquet systems: General formulation, special properties, and floquet
  topological defects},}\ }\href {\doibase 10.1103/PhysRevB.96.195303}
  {\bibfield  {journal} {\bibinfo  {journal} {Phys. Rev. B}\ }\textbf {\bibinfo
  {volume} {96}},\ \bibinfo {pages} {195303} (\bibinfo {year}
  {2017})}\BibitemShut {NoStop}%
\bibitem [{\citenamefont {Oka}\ and\ \citenamefont
  {Aoki}(2009{\natexlab{b}})}]{OKA}%
  \BibitemOpen
  \bibfield  {author} {\bibinfo {author} {\bibfnamefont {Takashi}\ \bibnamefont
  {Oka}}\ and\ \bibinfo {author} {\bibfnamefont {Hideo}\ \bibnamefont {Aoki}},\
  }\bibfield  {title} {\enquote {\bibinfo {title} {Photovoltaic hall effect in
  graphene},}\ }\href {\doibase 10.1103/PhysRevB.79.081406} {\bibfield
  {journal} {\bibinfo  {journal} {Phys. Rev. B}\ }\textbf {\bibinfo {volume}
  {79}},\ \bibinfo {pages} {081406(R)} (\bibinfo {year}
  {2009}{\natexlab{b}})}\BibitemShut {NoStop}%
\bibitem [{\citenamefont {Fulde}\ and\ \citenamefont
  {Ferrell}(1964)}]{Fulde_1964}%
  \BibitemOpen
  \bibfield  {author} {\bibinfo {author} {\bibfnamefont {Peter}\ \bibnamefont
  {Fulde}}\ and\ \bibinfo {author} {\bibfnamefont {Richard~A.}\ \bibnamefont
  {Ferrell}},\ }\bibfield  {title} {\enquote {\bibinfo {title}
  {Superconductivity in a strong spin-exchange field},}\ }\href {\doibase
  10.1103/PhysRev.135.A550} {\bibfield  {journal} {\bibinfo  {journal} {Phys.
  Rev.}\ }\textbf {\bibinfo {volume} {135}},\ \bibinfo {pages} {A550--A563}
  (\bibinfo {year} {1964})}\BibitemShut {NoStop}%
\bibitem [{\citenamefont {Larkin}\ and\ \citenamefont
  {Ovchinnikov}(1964)}]{Larkin_1964}%
  \BibitemOpen
  \bibfield  {author} {\bibinfo {author} {\bibfnamefont {A.~I.}\ \bibnamefont
  {Larkin}}\ and\ \bibinfo {author} {\bibfnamefont {Y.~N.}\ \bibnamefont
  {Ovchinnikov}},\ }\bibfield  {title} {\enquote {\bibinfo {title} {{Nonuniform
  state of superconductors}},}\ }\href@noop {} {\bibfield  {journal} {\bibinfo
  {journal} {Zh. Eksp. Teor. Fiz.}\ }\textbf {\bibinfo {volume} {47}},\
  \bibinfo {pages} {1136--1146} (\bibinfo {year} {1964})}\BibitemShut {NoStop}%
\bibitem [{\citenamefont {Bhowmik}\ and\ \citenamefont
  {Saha}(2025)}]{Sayak_2025}%
  \BibitemOpen
  \bibfield  {author} {\bibinfo {author} {\bibfnamefont {Sayak}\ \bibnamefont
  {Bhowmik}}\ and\ \bibinfo {author} {\bibfnamefont {Arijit}\ \bibnamefont
  {Saha}},\ }\bibfield  {title} {\enquote {\bibinfo {title} {Topological
  majorana zero modes and the superconducting diode effect driven by
  fulde-ferrell-larkin-ovchinnikov pairing in a helical shiba chain},}\ }\href
  {\doibase 10.1103/PhysRevB.111.L161402} {\bibfield  {journal} {\bibinfo
  {journal} {Phys. Rev. B}\ }\textbf {\bibinfo {volume} {111}},\ \bibinfo
  {pages} {L161402} (\bibinfo {year} {2025})}\BibitemShut {NoStop}%
\bibitem [{\citenamefont {Daido}\ and\ \citenamefont
  {Yanase}(2022)}]{Yanasediode}%
  \BibitemOpen
  \bibfield  {author} {\bibinfo {author} {\bibfnamefont {Akito}\ \bibnamefont
  {Daido}}\ and\ \bibinfo {author} {\bibfnamefont {Youichi}\ \bibnamefont
  {Yanase}},\ }\bibfield  {title} {\enquote {\bibinfo {title} {Superconducting
  diode effect and nonreciprocal transition lines},}\ }\href {\doibase
  10.1103/PhysRevB.106.205206} {\bibfield  {journal} {\bibinfo  {journal}
  {Phys. Rev. B}\ }\textbf {\bibinfo {volume} {106}},\ \bibinfo {pages}
  {205206} (\bibinfo {year} {2022})}\BibitemShut {NoStop}%
\bibitem [{\citenamefont {Daido}\ \emph {et~al.}(2022)\citenamefont {Daido},
  \citenamefont {Ikeda},\ and\ \citenamefont {Yanase}}]{Yanaseprl}%
  \BibitemOpen
  \bibfield  {author} {\bibinfo {author} {\bibfnamefont {Akito}\ \bibnamefont
  {Daido}}, \bibinfo {author} {\bibfnamefont {Yuhei}\ \bibnamefont {Ikeda}}, \
  and\ \bibinfo {author} {\bibfnamefont {Youichi}\ \bibnamefont {Yanase}},\
  }\bibfield  {title} {\enquote {\bibinfo {title} {Intrinsic superconducting
  diode effect},}\ }\href {\doibase 10.1103/PhysRevLett.128.037001} {\bibfield
  {journal} {\bibinfo  {journal} {Phys. Rev. Lett.}\ }\textbf {\bibinfo
  {volume} {128}},\ \bibinfo {pages} {037001} (\bibinfo {year}
  {2022})}\BibitemShut {NoStop}%
\bibitem [{\citenamefont {Nadeem}\ \emph {et~al.}(2023)\citenamefont {Nadeem},
  \citenamefont {Fuhrer},\ and\ \citenamefont
  {Wang}}]{nadeem2023superconducting}%
  \BibitemOpen
  \bibfield  {author} {\bibinfo {author} {\bibfnamefont {Muhammad}\
  \bibnamefont {Nadeem}}, \bibinfo {author} {\bibfnamefont {Michael~S}\
  \bibnamefont {Fuhrer}}, \ and\ \bibinfo {author} {\bibfnamefont {Xiaolin}\
  \bibnamefont {Wang}},\ }\bibfield  {title} {\enquote {\bibinfo {title} {The
  superconducting diode effect},}\ }\href {\doibase 10.1038/s42254-023-00632-w}
  {\bibfield  {journal} {\bibinfo  {journal} {Nature Reviews Physics}\ }\textbf
  {\bibinfo {volume} {5}},\ \bibinfo {pages} {558--577} (\bibinfo {year}
  {2023})}\BibitemShut {NoStop}%
\bibitem [{\citenamefont {Yuan}\ and\ \citenamefont {Fu}(2022)}]{LiangPNAS}%
  \BibitemOpen
  \bibfield  {author} {\bibinfo {author} {\bibfnamefont {Noah F.~Q.}\
  \bibnamefont {Yuan}}\ and\ \bibinfo {author} {\bibfnamefont {Liang}\
  \bibnamefont {Fu}},\ }\bibfield  {title} {\enquote {\bibinfo {title}
  {Supercurrent diode effect and finite-momentum superconductors},}\ }\href
  {\doibase 10.1073/pnas.2119548119} {\bibfield  {journal} {\bibinfo  {journal}
  {Proceedings of the National Academy of Sciences}\ }\textbf {\bibinfo
  {volume} {119}},\ \bibinfo {pages} {e2119548119} (\bibinfo {year}
  {2022})}\BibitemShut {NoStop}%
\bibitem [{\citenamefont {Ili\ifmmode~\acute{c}\else \'{c}\fi{}}\ and\
  \citenamefont {Bergeret}(2022)}]{SLlicprl}%
  \BibitemOpen
  \bibfield  {author} {\bibinfo {author} {\bibfnamefont {S.}~\bibnamefont
  {Ili\ifmmode~\acute{c}\else \'{c}\fi{}}}\ and\ \bibinfo {author}
  {\bibfnamefont {F.~S.}\ \bibnamefont {Bergeret}},\ }\bibfield  {title}
  {\enquote {\bibinfo {title} {Theory of the supercurrent diode effect in
  rashba superconductors with arbitrary disorder},}\ }\href {\doibase
  10.1103/PhysRevLett.128.177001} {\bibfield  {journal} {\bibinfo  {journal}
  {Phys. Rev. Lett.}\ }\textbf {\bibinfo {volume} {128}},\ \bibinfo {pages}
  {177001} (\bibinfo {year} {2022})}\BibitemShut {NoStop}%
\bibitem [{\citenamefont {de~Picoli}\ \emph {et~al.}(2023)\citenamefont
  {de~Picoli}, \citenamefont {Blood}, \citenamefont {Lyanda-Geller},\ and\
  \citenamefont {V\"ayrynen}}]{Picoli}%
  \BibitemOpen
  \bibfield  {author} {\bibinfo {author} {\bibfnamefont {Tatiana}\ \bibnamefont
  {de~Picoli}}, \bibinfo {author} {\bibfnamefont {Zane}\ \bibnamefont {Blood}},
  \bibinfo {author} {\bibfnamefont {Yuli}\ \bibnamefont {Lyanda-Geller}}, \
  and\ \bibinfo {author} {\bibfnamefont {Jukka~I.}\ \bibnamefont
  {V\"ayrynen}},\ }\bibfield  {title} {\enquote {\bibinfo {title}
  {Superconducting diode effect in quasi-one-dimensional systems},}\ }\href
  {\doibase 10.1103/PhysRevB.107.224518} {\bibfield  {journal} {\bibinfo
  {journal} {Phys. Rev. B}\ }\textbf {\bibinfo {volume} {107}},\ \bibinfo
  {pages} {224518} (\bibinfo {year} {2023})}\BibitemShut {NoStop}%
\bibitem [{\citenamefont {He}\ \emph {et~al.}(2022)\citenamefont {He},
  \citenamefont {Tanaka},\ and\ \citenamefont {Nagaosa}}]{Nagaosanjp}%
  \BibitemOpen
  \bibfield  {author} {\bibinfo {author} {\bibfnamefont {James~Jun}\
  \bibnamefont {He}}, \bibinfo {author} {\bibfnamefont {Yukio}\ \bibnamefont
  {Tanaka}}, \ and\ \bibinfo {author} {\bibfnamefont {Naoto}\ \bibnamefont
  {Nagaosa}},\ }\bibfield  {title} {\enquote {\bibinfo {title} {A
  phenomenological theory of superconductor diodes},}\ }\href {\doibase
  10.1088/1367-2630/ac6766} {\bibfield  {journal} {\bibinfo  {journal} {New
  Journal of Physics}\ }\textbf {\bibinfo {volume} {24}},\ \bibinfo {pages}
  {053014} (\bibinfo {year} {2022})}\BibitemShut {NoStop}%
\bibitem [{\citenamefont {Qu}\ \emph {et~al.}(2013{\natexlab{a}})\citenamefont
  {Qu}, \citenamefont {Zheng}, \citenamefont {Gong}, \citenamefont {Xu},
  \citenamefont {Mao}, \citenamefont {Zou}, \citenamefont {Guo},\ and\
  \citenamefont {Zhang}}]{Qu2013natcom}%
  \BibitemOpen
  \bibfield  {author} {\bibinfo {author} {\bibfnamefont {Chunlei}\ \bibnamefont
  {Qu}}, \bibinfo {author} {\bibfnamefont {Zhen}\ \bibnamefont {Zheng}},
  \bibinfo {author} {\bibfnamefont {Ming}\ \bibnamefont {Gong}}, \bibinfo
  {author} {\bibfnamefont {Yong}\ \bibnamefont {Xu}}, \bibinfo {author}
  {\bibfnamefont {Li}~\bibnamefont {Mao}}, \bibinfo {author} {\bibfnamefont
  {Xubo}\ \bibnamefont {Zou}}, \bibinfo {author} {\bibfnamefont {Guangcan}\
  \bibnamefont {Guo}}, \ and\ \bibinfo {author} {\bibfnamefont {Chuanwei}\
  \bibnamefont {Zhang}},\ }\bibfield  {title} {\enquote {\bibinfo {title}
  {Topological superfluids with finite-momentum pairing and majorana
  fermions},}\ }\href {\doibase 10.1038/ncomms3710} {\bibfield  {journal}
  {\bibinfo  {journal} {Nature Communications}\ }\textbf {\bibinfo {volume}
  {4}},\ \bibinfo {pages} {2710} (\bibinfo {year}
  {2013}{\natexlab{a}})}\BibitemShut {NoStop}%
\bibitem [{\citenamefont {Legg}\ \emph {et~al.}(2022)\citenamefont {Legg},
  \citenamefont {Loss},\ and\ \citenamefont {Klinovaja}}]{Lossdiode}%
  \BibitemOpen
  \bibfield  {author} {\bibinfo {author} {\bibfnamefont {Henry~F.}\
  \bibnamefont {Legg}}, \bibinfo {author} {\bibfnamefont {Daniel}\ \bibnamefont
  {Loss}}, \ and\ \bibinfo {author} {\bibfnamefont {Jelena}\ \bibnamefont
  {Klinovaja}},\ }\bibfield  {title} {\enquote {\bibinfo {title}
  {Superconducting diode effect due to magnetochiral anisotropy in topological
  insulators and rashba nanowires},}\ }\href {\doibase
  10.1103/PhysRevB.106.104501} {\bibfield  {journal} {\bibinfo  {journal}
  {Phys. Rev. B}\ }\textbf {\bibinfo {volume} {106}},\ \bibinfo {pages}
  {104501} (\bibinfo {year} {2022})}\BibitemShut {NoStop}%
\bibitem [{\citenamefont {Ruthvik}\ and\ \citenamefont
  {Nag}(2025)}]{ruthvik2025field}%
  \BibitemOpen
  \bibfield  {author} {\bibinfo {author} {\bibfnamefont {SVS}\ \bibnamefont
  {Ruthvik}}\ and\ \bibinfo {author} {\bibfnamefont {Tanay}\ \bibnamefont
  {Nag}},\ }\bibfield  {title} {\enquote {\bibinfo {title} {Field-free diode
  effects in one-dimensional superconductor: a complex interplay between
  fulde-ferrell pairing and altermagnetism},}\ }\href@noop {} {\bibfield
  {journal} {\bibinfo  {journal} {arXiv preprint arXiv:2512.01415}\ } (\bibinfo
  {year} {2025})}\BibitemShut {NoStop}%
\bibitem [{\citenamefont {Pal}\ \emph {et~al.}(2026)\citenamefont {Pal},
  \citenamefont {Mondal}, \citenamefont {Nag},\ and\ \citenamefont
  {Saha}}]{pal2025topological}%
  \BibitemOpen
  \bibfield  {author} {\bibinfo {author} {\bibfnamefont {Amartya}\ \bibnamefont
  {Pal}}, \bibinfo {author} {\bibfnamefont {Debashish}\ \bibnamefont {Mondal}},
  \bibinfo {author} {\bibfnamefont {Tanay}\ \bibnamefont {Nag}}, \ and\
  \bibinfo {author} {\bibfnamefont {Arijit}\ \bibnamefont {Saha}},\ }\bibfield
  {title} {\enquote {\bibinfo {title} {Topological superconductivity and
  superconducting diode effect mediated via unconventional magnet and ising
  spin-orbit coupling},}\ }\href {\doibase 10.1103/g4ry-j1xy} {\bibfield
  {journal} {\bibinfo  {journal} {Phys. Rev. B}\ }\textbf {\bibinfo {volume}
  {113}},\ \bibinfo {pages} {195409} (\bibinfo {year} {2026})}\BibitemShut
  {NoStop}%
\bibitem [{\citenamefont {Rikken}\ \emph {et~al.}(2001)\citenamefont {Rikken},
  \citenamefont {F\"olling},\ and\ \citenamefont
  {Wyder}}]{PhysRevLett.87.236602}%
  \BibitemOpen
  \bibfield  {author} {\bibinfo {author} {\bibfnamefont {G.~L. J.~A.}\
  \bibnamefont {Rikken}}, \bibinfo {author} {\bibfnamefont {J.}~\bibnamefont
  {F\"olling}}, \ and\ \bibinfo {author} {\bibfnamefont {P.}~\bibnamefont
  {Wyder}},\ }\bibfield  {title} {\enquote {\bibinfo {title} {Electrical
  magnetochiral anisotropy},}\ }\href {\doibase 10.1103/PhysRevLett.87.236602}
  {\bibfield  {journal} {\bibinfo  {journal} {Phys. Rev. Lett.}\ }\textbf
  {\bibinfo {volume} {87}},\ \bibinfo {pages} {236602} (\bibinfo {year}
  {2001})}\BibitemShut {NoStop}%
\bibitem [{\citenamefont {Yokouchi}\ \emph {et~al.}(2023)\citenamefont
  {Yokouchi}, \citenamefont {Ikeda}, \citenamefont {Morimoto},\ and\
  \citenamefont {Shiomi}}]{PhysRevLett.130.136301}%
  \BibitemOpen
  \bibfield  {author} {\bibinfo {author} {\bibfnamefont {Tomoyuki}\
  \bibnamefont {Yokouchi}}, \bibinfo {author} {\bibfnamefont {Yuya}\
  \bibnamefont {Ikeda}}, \bibinfo {author} {\bibfnamefont {Takahiro}\
  \bibnamefont {Morimoto}}, \ and\ \bibinfo {author} {\bibfnamefont {Yuki}\
  \bibnamefont {Shiomi}},\ }\bibfield  {title} {\enquote {\bibinfo {title}
  {Giant magnetochiral anisotropy in weyl semimetal ${\mathrm{wte}}_{2}$
  induced by diverging berry curvature},}\ }\href {\doibase
  10.1103/PhysRevLett.130.136301} {\bibfield  {journal} {\bibinfo  {journal}
  {Phys. Rev. Lett.}\ }\textbf {\bibinfo {volume} {130}},\ \bibinfo {pages}
  {136301} (\bibinfo {year} {2023})}\BibitemShut {NoStop}%
\bibitem [{\citenamefont {Rikken}\ and\ \citenamefont
  {Avarvari}(2019)}]{PhysRevB.99.245153}%
  \BibitemOpen
  \bibfield  {author} {\bibinfo {author} {\bibfnamefont {G.~L. J.~A.}\
  \bibnamefont {Rikken}}\ and\ \bibinfo {author} {\bibfnamefont
  {N.}~\bibnamefont {Avarvari}},\ }\bibfield  {title} {\enquote {\bibinfo
  {title} {Strong electrical magnetochiral anisotropy in tellurium},}\ }\href
  {\doibase 10.1103/PhysRevB.99.245153} {\bibfield  {journal} {\bibinfo
  {journal} {Phys. Rev. B}\ }\textbf {\bibinfo {volume} {99}},\ \bibinfo
  {pages} {245153} (\bibinfo {year} {2019})}\BibitemShut {NoStop}%
\bibitem [{\citenamefont {Ando}\ \emph {et~al.}(2020)\citenamefont {Ando},
  \citenamefont {Miyasaka}, \citenamefont {Li}, \citenamefont {Ishizuka},
  \citenamefont {Arakawa}, \citenamefont {Shiota}, \citenamefont {Moriyama},
  \citenamefont {Yanase},\ and\ \citenamefont {Ono}}]{ando2020observation}%
  \BibitemOpen
  \bibfield  {author} {\bibinfo {author} {\bibfnamefont {Fuyuki}\ \bibnamefont
  {Ando}}, \bibinfo {author} {\bibfnamefont {Yuta}\ \bibnamefont {Miyasaka}},
  \bibinfo {author} {\bibfnamefont {Tian}\ \bibnamefont {Li}}, \bibinfo
  {author} {\bibfnamefont {Jun}\ \bibnamefont {Ishizuka}}, \bibinfo {author}
  {\bibfnamefont {Tomonori}\ \bibnamefont {Arakawa}}, \bibinfo {author}
  {\bibfnamefont {Yoichi}\ \bibnamefont {Shiota}}, \bibinfo {author}
  {\bibfnamefont {Takahiro}\ \bibnamefont {Moriyama}}, \bibinfo {author}
  {\bibfnamefont {Youichi}\ \bibnamefont {Yanase}}, \ and\ \bibinfo {author}
  {\bibfnamefont {Teruo}\ \bibnamefont {Ono}},\ }\bibfield  {title} {\enquote
  {\bibinfo {title} {Observation of superconducting diode effect},}\ }\href
  {\doibase 10.1038/s41586-020-2590-4} {\bibfield  {journal} {\bibinfo
  {journal} {Nature}\ }\textbf {\bibinfo {volume} {584}},\ \bibinfo {pages}
  {373--376} (\bibinfo {year} {2020})}\BibitemShut {NoStop}%
\bibitem [{\citenamefont {Nakamura}\ \emph {et~al.}(2024)\citenamefont
  {Nakamura}, \citenamefont {Daido},\ and\ \citenamefont
  {Yanase}}]{PhysRevB.109.094501}%
  \BibitemOpen
  \bibfield  {author} {\bibinfo {author} {\bibfnamefont {Kyohei}\ \bibnamefont
  {Nakamura}}, \bibinfo {author} {\bibfnamefont {Akito}\ \bibnamefont {Daido}},
  \ and\ \bibinfo {author} {\bibfnamefont {Youichi}\ \bibnamefont {Yanase}},\
  }\bibfield  {title} {\enquote {\bibinfo {title} {Orbital effect on the
  intrinsic superconducting diode effect},}\ }\href {\doibase
  10.1103/PhysRevB.109.094501} {\bibfield  {journal} {\bibinfo  {journal}
  {Phys. Rev. B}\ }\textbf {\bibinfo {volume} {109}},\ \bibinfo {pages}
  {094501} (\bibinfo {year} {2024})}\BibitemShut {NoStop}%
\bibitem [{\citenamefont {Qu}\ \emph {et~al.}(2013{\natexlab{b}})\citenamefont
  {Qu}, \citenamefont {Zheng}, \citenamefont {Gong}, \citenamefont {Xu},
  \citenamefont {Mao}, \citenamefont {Zou}, \citenamefont {Guo},\ and\
  \citenamefont {Zhang}}]{qu2013topological}%
  \BibitemOpen
  \bibfield  {author} {\bibinfo {author} {\bibfnamefont {Chunlei}\ \bibnamefont
  {Qu}}, \bibinfo {author} {\bibfnamefont {Zhen}\ \bibnamefont {Zheng}},
  \bibinfo {author} {\bibfnamefont {Ming}\ \bibnamefont {Gong}}, \bibinfo
  {author} {\bibfnamefont {Yong}\ \bibnamefont {Xu}}, \bibinfo {author}
  {\bibfnamefont {Li}~\bibnamefont {Mao}}, \bibinfo {author} {\bibfnamefont
  {Xubo}\ \bibnamefont {Zou}}, \bibinfo {author} {\bibfnamefont {Guangcan}\
  \bibnamefont {Guo}}, \ and\ \bibinfo {author} {\bibfnamefont {Chuanwei}\
  \bibnamefont {Zhang}},\ }\bibfield  {title} {\enquote {\bibinfo {title}
  {Topological superfluids with finite-momentum pairing and majorana
  fermions},}\ }\href {\doibase 10.1038/ncomms3710} {\bibfield  {journal}
  {\bibinfo  {journal} {Nature communications}\ }\textbf {\bibinfo {volume}
  {4}},\ \bibinfo {pages} {2710} (\bibinfo {year}
  {2013}{\natexlab{b}})}\BibitemShut {NoStop}%
\bibitem [{\citenamefont {Hu}\ \emph {et~al.}(2019)\citenamefont {Hu},
  \citenamefont {Liu},\ and\ \citenamefont {Zhang}}]{hu2019topological}%
  \BibitemOpen
  \bibfield  {author} {\bibinfo {author} {\bibfnamefont {Lun-Hui}\ \bibnamefont
  {Hu}}, \bibinfo {author} {\bibfnamefont {Chao-Xing}\ \bibnamefont {Liu}}, \
  and\ \bibinfo {author} {\bibfnamefont {Fu-Chun}\ \bibnamefont {Zhang}},\
  }\bibfield  {title} {\enquote {\bibinfo {title} {Topological
  larkin-ovchinnikov phase and majorana zero mode chain in bilayer
  superconducting topological insulator films},}\ }\href {\doibase
  10.1038/s42005-019-0126-8} {\bibfield  {journal} {\bibinfo  {journal}
  {Communications Physics}\ }\textbf {\bibinfo {volume} {2}},\ \bibinfo {pages}
  {25} (\bibinfo {year} {2019})}\BibitemShut {NoStop}%
\bibitem [{\citenamefont {Zhu}(2016)}]{Zhu_2016}%
  \BibitemOpen
  \bibfield  {author} {\bibinfo {author} {\bibfnamefont {Jian-Xin}\
  \bibnamefont {Zhu}},\ }\href@noop {} {\emph {\bibinfo {title} {Bogoliubov-de
  Gennes method and its applications}}},\ Vol.\ \bibinfo {volume} {924}\
  (\bibinfo  {publisher} {Springer},\ \bibinfo {year} {2016})\BibitemShut
  {NoStop}%
\bibitem [{sup()}]{supp}%
  \BibitemOpen
  \href@noop {} {}\bibinfo {note} {See the Supplementary Material at
  XXXXXXXXXXX for the detailed discussion on Derivation of the Floquet Spectral
  Properties, Reciprocal Floquet Spectra and Supercurrent Characteristics for
  $B_y=0$}\BibitemShut {NoStop}%
\bibitem [{\citenamefont {Resta}(1998)}]{Resta}%
  \BibitemOpen
  \bibfield  {author} {\bibinfo {author} {\bibfnamefont {Raffaele}\
  \bibnamefont {Resta}},\ }\bibfield  {title} {\enquote {\bibinfo {title}
  {Quantum-mechanical position operator in extended systems},}\ }\href
  {\doibase 10.1103/PhysRevLett.80.1800} {\bibfield  {journal} {\bibinfo
  {journal} {Phys. Rev. Lett.}\ }\textbf {\bibinfo {volume} {80}},\ \bibinfo
  {pages} {1800--1803} (\bibinfo {year} {1998})}\BibitemShut {NoStop}%
\bibitem [{\citenamefont {Wheeler}\ \emph {et~al.}(2019)\citenamefont
  {Wheeler}, \citenamefont {Wagner},\ and\ \citenamefont {Hughes}}]{Wheeler}%
  \BibitemOpen
  \bibfield  {author} {\bibinfo {author} {\bibfnamefont {William~A.}\
  \bibnamefont {Wheeler}}, \bibinfo {author} {\bibfnamefont {Lucas~K.}\
  \bibnamefont {Wagner}}, \ and\ \bibinfo {author} {\bibfnamefont {Taylor~L.}\
  \bibnamefont {Hughes}},\ }\bibfield  {title} {\enquote {\bibinfo {title}
  {Many-body electric multipole operators in extended systems},}\ }\href
  {\doibase 10.1103/PhysRevB.100.245135} {\bibfield  {journal} {\bibinfo
  {journal} {Phys. Rev. B}\ }\textbf {\bibinfo {volume} {100}},\ \bibinfo
  {pages} {245135} (\bibinfo {year} {2019})}\BibitemShut {NoStop}%
\bibitem [{\citenamefont {Kang}\ \emph {et~al.}(2019)\citenamefont {Kang},
  \citenamefont {Shiozaki},\ and\ \citenamefont {Cho}}]{KangPRB}%
  \BibitemOpen
  \bibfield  {author} {\bibinfo {author} {\bibfnamefont {Byungmin}\
  \bibnamefont {Kang}}, \bibinfo {author} {\bibfnamefont {Ken}\ \bibnamefont
  {Shiozaki}}, \ and\ \bibinfo {author} {\bibfnamefont {Gil~Young}\
  \bibnamefont {Cho}},\ }\bibfield  {title} {\enquote {\bibinfo {title}
  {Many-body order parameters for multipoles in solids},}\ }\href {\doibase
  10.1103/PhysRevB.100.245134} {\bibfield  {journal} {\bibinfo  {journal}
  {Phys. Rev. B}\ }\textbf {\bibinfo {volume} {100}},\ \bibinfo {pages}
  {245134} (\bibinfo {year} {2019})}\BibitemShut {NoStop}%
\bibitem [{\citenamefont {Bhowmik}\ \emph {et~al.}(2025)\citenamefont
  {Bhowmik}, \citenamefont {Samanta}, \citenamefont {Nandy}, \citenamefont
  {Saha},\ and\ \citenamefont {Ghosh}}]{bhowmik_2025}%
  \BibitemOpen
  \bibfield  {author} {\bibinfo {author} {\bibfnamefont {Sayak}\ \bibnamefont
  {Bhowmik}}, \bibinfo {author} {\bibfnamefont {Dibyendu}\ \bibnamefont
  {Samanta}}, \bibinfo {author} {\bibfnamefont {Ashis~K}\ \bibnamefont
  {Nandy}}, \bibinfo {author} {\bibfnamefont {Arijit}\ \bibnamefont {Saha}}, \
  and\ \bibinfo {author} {\bibfnamefont {Sudeep~Kumar}\ \bibnamefont {Ghosh}},\
  }\bibfield  {title} {\enquote {\bibinfo {title} {Optimizing one dimensional
  superconducting diodes: interplay of rashba spin-orbit coupling and magnetic
  fields},}\ }\href {https://doi.org/10.1038/s42005-025-02044-x} {\bibfield
  {journal} {\bibinfo  {journal} {Communications Physics}\ }\textbf {\bibinfo
  {volume} {8}},\ \bibinfo {pages} {260} (\bibinfo {year} {2025})}\BibitemShut
  {NoStop}%
\bibitem [{\citenamefont {Diez-Merida}\ \emph {et~al.}(2023)\citenamefont
  {Diez-Merida}, \citenamefont {D{\'\i}ez-Carl{\'o}n}, \citenamefont {Yang},
  \citenamefont {Xie}, \citenamefont {Gao}, \citenamefont {Senior},
  \citenamefont {Watanabe}, \citenamefont {Taniguchi}, \citenamefont {Lu},
  \citenamefont {Higginbotham} \emph {et~al.}}]{Jaime_2023}%
  \BibitemOpen
  \bibfield  {author} {\bibinfo {author} {\bibfnamefont {Jaime}\ \bibnamefont
  {Diez-Merida}}, \bibinfo {author} {\bibfnamefont {Andr{\'e}s}\ \bibnamefont
  {D{\'\i}ez-Carl{\'o}n}}, \bibinfo {author} {\bibfnamefont {SY}~\bibnamefont
  {Yang}}, \bibinfo {author} {\bibfnamefont {Y-M}\ \bibnamefont {Xie}},
  \bibinfo {author} {\bibfnamefont {X-J}\ \bibnamefont {Gao}}, \bibinfo
  {author} {\bibfnamefont {Jorden}\ \bibnamefont {Senior}}, \bibinfo {author}
  {\bibfnamefont {K}~\bibnamefont {Watanabe}}, \bibinfo {author} {\bibfnamefont
  {T}~\bibnamefont {Taniguchi}}, \bibinfo {author} {\bibfnamefont
  {X}~\bibnamefont {Lu}}, \bibinfo {author} {\bibfnamefont {Andrew~P}\
  \bibnamefont {Higginbotham}},  \emph {et~al.},\ }\bibfield  {title} {\enquote
  {\bibinfo {title} {Symmetry-broken josephson junctions and superconducting
  diodes in magic-angle twisted bilayer graphene},}\ }\href
  {https://doi.org/10.1038/s41467-023-38005-7} {\bibfield  {journal} {\bibinfo
  {journal} {Nature Communications}\ }\textbf {\bibinfo {volume} {14}},\
  \bibinfo {pages} {2396} (\bibinfo {year} {2023})}\BibitemShut {NoStop}%
\bibitem [{\citenamefont {Wang}\ \emph {et~al.}(2013)\citenamefont {Wang},
  \citenamefont {Steinberg}, \citenamefont {Jarillo-Herrero},\ and\
  \citenamefont {Gedik}}]{wang2013observation}%
  \BibitemOpen
  \bibfield  {author} {\bibinfo {author} {\bibfnamefont {YH}~\bibnamefont
  {Wang}}, \bibinfo {author} {\bibfnamefont {Hadar}\ \bibnamefont {Steinberg}},
  \bibinfo {author} {\bibfnamefont {Pablo}\ \bibnamefont {Jarillo-Herrero}}, \
  and\ \bibinfo {author} {\bibfnamefont {Nuh}\ \bibnamefont {Gedik}},\
  }\bibfield  {title} {\enquote {\bibinfo {title} {Observation of floquet-bloch
  states on the surface of a topological insulator},}\ }\href {\doibase
  10.1126/science.1239834} {\bibfield  {journal} {\bibinfo  {journal}
  {Science}\ }\textbf {\bibinfo {volume} {342}},\ \bibinfo {pages} {453--457}
  (\bibinfo {year} {2013})}\BibitemShut {NoStop}%
\bibitem [{\citenamefont {Mahmood}\ \emph {et~al.}(2016)\citenamefont
  {Mahmood}, \citenamefont {Chan}, \citenamefont {Alpichshev}, \citenamefont
  {Gardner}, \citenamefont {Lee}, \citenamefont {Lee},\ and\ \citenamefont
  {Gedik}}]{Mahmood2016}%
  \BibitemOpen
  \bibfield  {author} {\bibinfo {author} {\bibfnamefont {Fahad}\ \bibnamefont
  {Mahmood}}, \bibinfo {author} {\bibfnamefont {Ching-Kit}\ \bibnamefont
  {Chan}}, \bibinfo {author} {\bibfnamefont {Zhanybek}\ \bibnamefont
  {Alpichshev}}, \bibinfo {author} {\bibfnamefont {Dillon}\ \bibnamefont
  {Gardner}}, \bibinfo {author} {\bibfnamefont {Young}\ \bibnamefont {Lee}},
  \bibinfo {author} {\bibfnamefont {Patrick~A.}\ \bibnamefont {Lee}}, \ and\
  \bibinfo {author} {\bibfnamefont {Nuh}\ \bibnamefont {Gedik}},\ }\bibfield
  {title} {\enquote {\bibinfo {title} {Selective scattering between
  floquet--bloch and volkov states in a topological insulator},}\ }\href
  {\doibase 10.1038/nphys3609} {\bibfield  {journal} {\bibinfo  {journal}
  {Nature Physics}\ }\textbf {\bibinfo {volume} {12}},\ \bibinfo {pages}
  {306--310} (\bibinfo {year} {2016})}\BibitemShut {NoStop}%
\bibitem [{\citenamefont {McIver}\ \emph {et~al.}(2020)\citenamefont {McIver},
  \citenamefont {Schulte}, \citenamefont {Stein}, \citenamefont {Matsuyama},
  \citenamefont {Jotzu}, \citenamefont {Meier},\ and\ \citenamefont
  {Cavalleri}}]{McIver2020}%
  \BibitemOpen
  \bibfield  {author} {\bibinfo {author} {\bibfnamefont {J.~W.}\ \bibnamefont
  {McIver}}, \bibinfo {author} {\bibfnamefont {B.}~\bibnamefont {Schulte}},
  \bibinfo {author} {\bibfnamefont {F.-U.}\ \bibnamefont {Stein}}, \bibinfo
  {author} {\bibfnamefont {T.}~\bibnamefont {Matsuyama}}, \bibinfo {author}
  {\bibfnamefont {G.}~\bibnamefont {Jotzu}}, \bibinfo {author} {\bibfnamefont
  {G.}~\bibnamefont {Meier}}, \ and\ \bibinfo {author} {\bibfnamefont
  {A.}~\bibnamefont {Cavalleri}},\ }\bibfield  {title} {\enquote {\bibinfo
  {title} {Light-induced anomalous hall effect in graphene},}\ }\href {\doibase
  10.1038/s41567-019-0698-y} {\bibfield  {journal} {\bibinfo  {journal} {Nature
  Physics}\ }\textbf {\bibinfo {volume} {16}},\ \bibinfo {pages} {38--41}
  (\bibinfo {year} {2020})}\BibitemShut {NoStop}%
\bibitem [{\citenamefont {Basov}\ \emph {et~al.}(2017)\citenamefont {Basov},
  \citenamefont {Averitt},\ and\ \citenamefont {Hsieh}}]{Basov2017}%
  \BibitemOpen
  \bibfield  {author} {\bibinfo {author} {\bibfnamefont {D.~N.}\ \bibnamefont
  {Basov}}, \bibinfo {author} {\bibfnamefont {R.~D.}\ \bibnamefont {Averitt}},
  \ and\ \bibinfo {author} {\bibfnamefont {D.}~\bibnamefont {Hsieh}},\
  }\bibfield  {title} {\enquote {\bibinfo {title} {Towards properties on demand
  in quantum materials},}\ }\href {\doibase 10.1038/nmat5017} {\bibfield
  {journal} {\bibinfo  {journal} {Nature Materials}\ }\textbf {\bibinfo
  {volume} {16}},\ \bibinfo {pages} {1077--1088} (\bibinfo {year}
  {2017})}\BibitemShut {NoStop}%
\bibitem [{\citenamefont {Lin}\ \emph {et~al.}(2022)\citenamefont {Lin},
  \citenamefont {Siriviboon}, \citenamefont {Scammell}, \citenamefont {Liu},
  \citenamefont {Rhodes}, \citenamefont {Watanabe}, \citenamefont {Taniguchi},
  \citenamefont {Hone}, \citenamefont {Scheurer},\ and\ \citenamefont
  {Li}}]{lin2022zero}%
  \BibitemOpen
  \bibfield  {author} {\bibinfo {author} {\bibfnamefont {Jiang-Xiazi}\
  \bibnamefont {Lin}}, \bibinfo {author} {\bibfnamefont {Phum}\ \bibnamefont
  {Siriviboon}}, \bibinfo {author} {\bibfnamefont {Harley~D}\ \bibnamefont
  {Scammell}}, \bibinfo {author} {\bibfnamefont {Song}\ \bibnamefont {Liu}},
  \bibinfo {author} {\bibfnamefont {Daniel}\ \bibnamefont {Rhodes}}, \bibinfo
  {author} {\bibfnamefont {K}~\bibnamefont {Watanabe}}, \bibinfo {author}
  {\bibfnamefont {T}~\bibnamefont {Taniguchi}}, \bibinfo {author}
  {\bibfnamefont {James}\ \bibnamefont {Hone}}, \bibinfo {author}
  {\bibfnamefont {Mathias~S}\ \bibnamefont {Scheurer}}, \ and\ \bibinfo
  {author} {\bibfnamefont {JIA}\ \bibnamefont {Li}},\ }\bibfield  {title}
  {\enquote {\bibinfo {title} {Zero-field superconducting diode effect in
  small-twist-angle trilayer graphene},}\ }\href {\doibase
  10.1038/s41567-022-01700-1} {\bibfield  {journal} {\bibinfo  {journal}
  {Nature Physics}\ }\textbf {\bibinfo {volume} {18}},\ \bibinfo {pages}
  {1221--1227} (\bibinfo {year} {2022})}\BibitemShut {NoStop}%
\end{thebibliography}%

\clearpage

\renewcommand{\theequation}{S\arabic{equation}}
\renewcommand{\thefigure}{S\arabic{figure}}
\renewcommand{\bibnumfmt}[1]{[S#1]}
\renewcommand{\citenumfont}[1]{S#1}

\newcommand{\dr}{\ensuremath{\mathbf{r}}}
\newpage
\onecolumngrid
\begin{center}

	{
		\fontsize{12}{12}
		\selectfont
		\textbf{Supplemental material for ``Superconducting diode effect via Floquet topological Fulde-Ferrell phase in driven Rashba nanowire''
			\\[5mm]}
	}
	\normalsize Sayak Bhowmik$^{1,2}$, Arijit Saha$^{1,2}$ and Tanay Nag$^{3}$\\
	\vspace{2mm}
	{\small $^1$\textit{Institute of Physics, Sachivalaya Marg, Bhubaneswar-751005, India}\\[0.5mm]}
	{\small $^2$\textit{Homi Bhabha National Institute, Training School Complex, Anushakti Nagar, Mumbai 400094, India}\\[0.5mm]}
	{\small $^2$\textit{Department of Physics, BITS Pilani-Hyderabad Campus, Telangana 500078, India}\\[0.5mm]}
	
\end{center}

\renewcommand{\thesection}{S\arabic{section}}

\tableofcontents 

\section{Derivation of the Floquet Spectral Properties  \label{sec:sec_1}}
In this section we provide the detailed derivation of the generic relation for the Floquet quasienergy dependence on the externally tunable driving strength ${J_1}^{'}$ as mentioned in the main text: 
\begin{equation}
	E_n(J^{\prime}_1+2\pi)=E_n(J^{\prime}_1)+\pi \ .
	\label{Eq1}
\end{equation} 

Given the three step  driving protocol consisting of external Zeeman fields $(B_x,B_y)$, the Floquet operator as described in Eq.~(3) of the main text reads as 
\begin{equation}
	\begin{aligned}
		U(T) &= e^{-i J_1 H_1 T/4} \, e^{-i J_0 H_0(k) T/2} \, e^{-i J_1 H_1 T/4} \\
		&= e^{-i J_1' H_1 /4} \, e^{-i J_0' H_0(k) /2} \, e^{-i J_1' H_1 /4}\ ,
	\end{aligned}
\end{equation}
where, $H_0$ represents the static Hamiltonian [see Eq.~(1) of the main text] and  $ H_1 =\sum_{n,s,s'}  c_{n,s}^\dagger \, (\mathbf{B} \cdot \boldsymbol{\sigma})_{ss'} \, c_{n,s'}$ with the Zeeman field components $\mathbf{B}=(B_x, B_y)$ and $T$ being the time period of the Floquet drive with $\mathbf{B}$ being a unit vector, as considered throughout or analysis. The Hamiltonian $H_1$ clearly takes the diagonal form in the particle-hole sector 
($\tau$ space). Then further dropping the summation over sites $n$, the $H_1$ kernel takes the form 
\begin{equation}
	{H_1} = \tau_0 (B_x \sigma_x+ B_y \sigma_y)\ .  
\end{equation}    
For \( |B| = 1 \), the Hamiltonian \( H_1 \) satisfies the relation \( H_1^2 = I \), with \( I \) being the identity matrix. As a result, one obtains
\begin{equation}
	\begin{aligned}
		e^{i J_1' H_1/4}
		&= \cos\left(\frac{J_1'}{4}\right) I
		+ i \sin\left(\frac{J_1'}{4}\right) H_1\ , \\
		e^{i (J_1'+2\pi) H_1/4}
		&= \cos\left(\frac{J_1'}{4}+\frac{\pi}{2}\right) I
		+ i \sin\left(\frac{J_1'}{4}+\frac{\pi}{2}\right) H_1
		= -\sin\left(\frac{J_1'}{4}\right) I
		+ i \cos\left(\frac{J_1'}{4}\right) H_1\ , \\
		e^{i (J_1'+2\pi) H_1/4}&= iH_1 e^{i J_1' H_1/4}\ .
	\end{aligned}
\end{equation}

Following this procedure, the Floquet operator $U(J_1'+2\pi)$ takes the form 
\begin{equation}
	\begin{aligned}
		U(J_1'+2\pi)
		&=
		(iH_1)\, e^{i J_1' H_1/4}
		\, e^{-i J_0' H_0(k)/2}
		\, e^{i J_1' H_1/4}\, (iH_1)\ , \\
		U(J_1'+2\pi) &= -\,H_1 U(J_1') H_1 \ .
	\end{aligned}
\end{equation}
Since, $H_1$ being an unitary operator, the  quasi-eigen spectra for the Floquet operator $U(J_1'+2\pi)$ can be written as 
\begin{equation}
	\begin{aligned}
		e^{-i E_n (J_1'+2\pi)}
		=
		-\, e^{-i E_n (J_1')}
		=
		e^{-i \left[E_n(J_1')+\pi\right]} \ .
	\end{aligned}
\end{equation}	
Hence, the quasienergy spectrum corresponding to the three step driving protocol satisfies the relation \(E_n(J_1'+2\pi)=E_n(J_1')+\pi\). Notably, this result is generic to the specific structure of \(H_1\) and remains independent of the explicit form of \(H_0\).

\section{Reciprocal Floquet Spectra and Supercurrent Characteristics for \texorpdfstring{$\boldsymbol{B_y=0}$}{By=0}}

\label{sec:sec_2}
\begin{figure*}[h]
	\centering \includegraphics[width=\columnwidth]{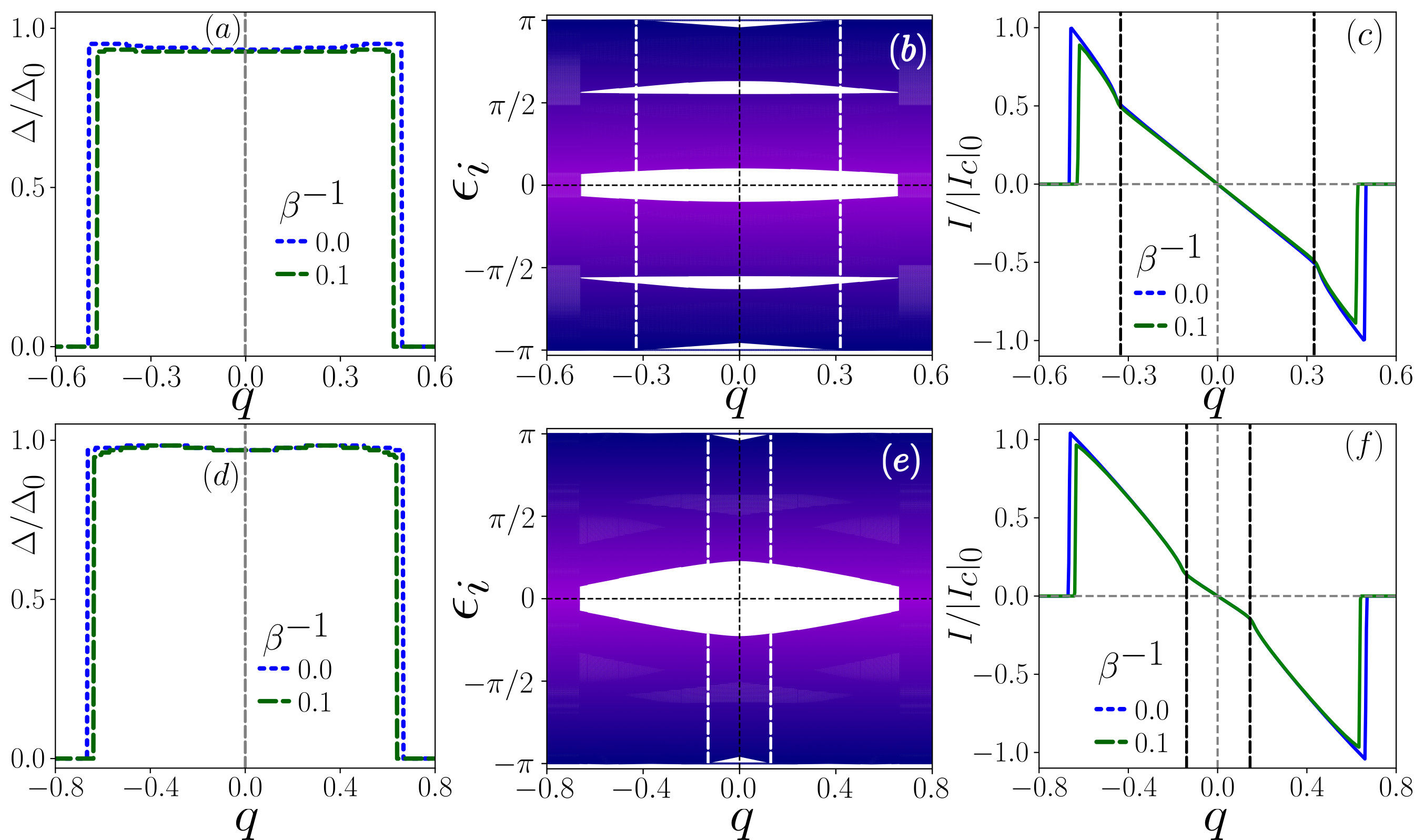}
	\caption{\textbf{Results for $\boldsymbol{B_y=0$}:} Numerical results obtained within the self-consistent mean-field framework are shown for the Floquet lattice Hamiltonian $\mathcal{H}_F$ [Eq.~(3) of main text] on a finite size lattice consisting of $300$ sites with $B_y=0$. The superconducting order parameter profile $\Delta(q)$, the Floquet quasienergy spectrum and the corresponding supercurrent profile $I(q)$ are depicted as a function of Cooper pair momentum $q$ for $(J_1', J_0')=(0.6\pi, 0.46\pi)$ in panels (a), (b), and (c) respectively and for $(J_1', J_0')=(0.4\pi, 0.6\pi)$ in panels (d), (e), and (f) respectively. The $\Delta(q)$ profile and the corresponding $I(q)$ profile are shown for two different temperatures $\beta^{-1}=(0.0,0.1)$. The dotted vertical lines indicate the topological phase boundaries for $q$ exhibiting $\pi$-FMMs and their signatures in supercurrent. Other model parameters are chosen as  $(t, \alpha)= 1{J_0}^{'}, U=2.3, \mu/J_0'=1.2$.}
	\label{fig:SFig1}
\end{figure*}

Here, we present the spectral properties and the corresponding supercurrent signatures for the \(B_y=0\) case within the self-consistent treatment of the superconducting order parameter. As shown in Fig.~\ref{fig:SFig1}(a),(d), the superconducting gap profile \(\Delta(q)\) remains symmetric with respect to \(q\) and, unlike the finite-\(B_y\) case discussed in the main text, exhibits no abrupt suppression in the positive-\(q\) sector. Consequently, the quasi-energy spectrum also preserves the symmetry under \(q \rightarrow -q\) and does not exhibit topological transitions in the positive-\(q\) regime [see Figs.~\ref{fig:SFig1}(b),(e)]. Nevertheless, the spectrum hosts a Floquet TSC phase supporting \(\pi\)-FMMs over a symmetric finite interval of \(q\) 
centered around \(q=0\).

As expected, the supercurrent is reciprocal, satisfying \(I(q)=I(-q)\), with the superconducting ground state occurring at zero Cooper-pair momentum, \(q_0=0\), i.e., \(I(q=0)=0\), thereby ruling out the possibility of finite-momentum FF pairing. However, the variation of the supercurrent with \(q\) still clearly identifies the topological regime through a distinct change in slope across the \(\pi\)-FMM phase, consistent with the behavior discussed in the main text. 
Therefore, this feature establishes the pivotal role of finite \(B_y\) in generating nonreciprocal superconducting response, enabling supercurrent controlled switching between topological superconducting phases hosting \(0\)- and \(\pi\)-FMMs, and facilitating the emergence of superconducting diode effect (SDE). 

\begin{figure*}[h]
	\centering \includegraphics[width=1.02\columnwidth]{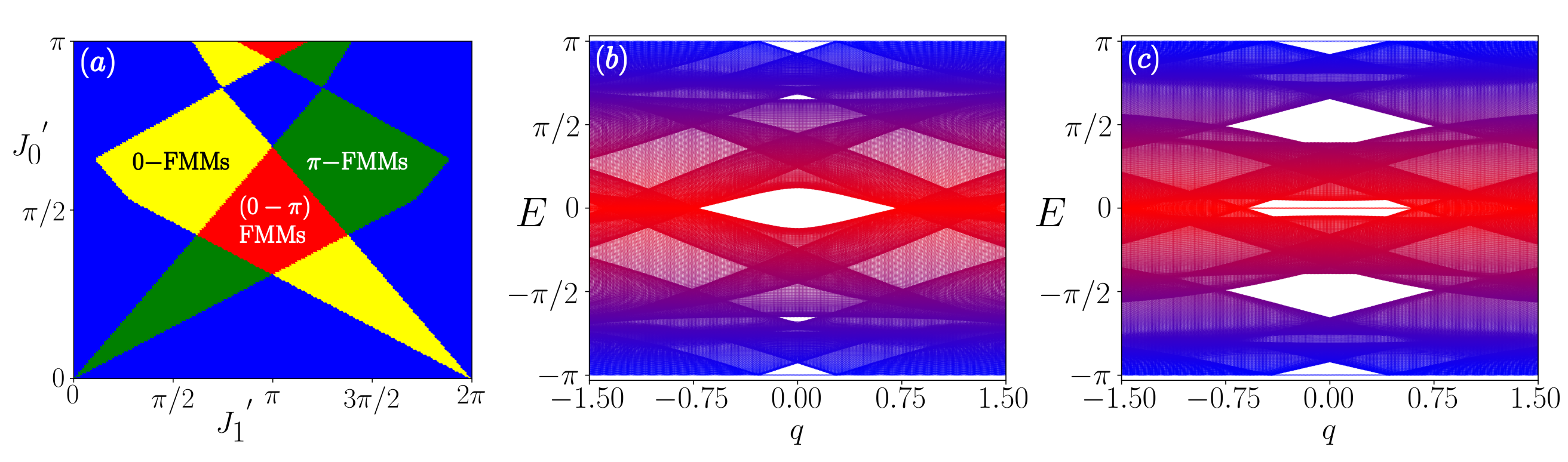}
	\caption{\textbf{Results for non self-consistent approach:} Numerical results obtained within the non-self-consistent framework considering a constant superconducting order parameter $\Delta$ are shown. (a) Topological phase diagram is displayed in the ${J_1}^{'}$–${J_0}^{'}$ plane for $q=0$, exhibiting distinct topological regimes hosting $0$-FMMs (yellow), $\pi$-FMMs (green), coexisting $0$–$\pi$ FMMs (red) and blue regime indicates non-topological phase. (b) Floquet quasi-energy spectrum is depicted as a function of finite Cooper-pair momentum $q$ for the parameter set $({J_1}^{'}$,${J_0}^{'})=(0.6\pi, 0.6\pi)$ and for $({J_1}^{'}$,${J_0}^{'})=(0.8\pi, 0.7\pi)$ in panel (c). 
		Other model parameters are chosen as: $(\mu, \alpha, t,\Delta)=1J_0$.}
	\label{fig:SFig2}
\end{figure*}

Then, we additionally present the corresponding results for $B_y=0$, obtained within the non-self-consistent approach, where the superconducting order parameter $\Delta$ is considered to be constant throughout the variation of the model parameters. Since the non-self-consistent treatment does not introduce additional qualitative features, we mainly focus here on the spectral and topological signatures. We demonstrate the topological phase diagram in the ${J_1}^{'}$–${J_0}^{'}$ plane for $q=0$ ($s$-wave) in Fig.~\ref{fig:SFig2}(a), exhibiting three distinct Floquet topological regimes hosting $0$-FMMs (yellow), $\pi$-FMMs (green), and coexisting $0$–$\pi$ FMMs (red). The quasienergy spectrum is depicted as a function of $q$ in Figs.~\ref{fig:SFig2}(b),(c). The spectrum remains symmetric about $q=0$, thereby preserving the topological character of the parent phase at $q=0$ upon extending symmetrically across the positive and negative $q$ sectors. In particular, Fig.~\ref{fig:SFig2}(b), corresponding to a representative point from the Floquet TSC phase hosting $\pi$-FMM only in Fig.~\ref{fig:SFig2}(a), retains the $\pi$-mode character symmetrically across the $\pm q$ sectors. Similarly, Fig.~\ref{fig:SFig2}(c), chosen from the coexisting $0$–$\pi$ FMMs regime of Fig.~\ref{fig:SFig2}(a), preserves the simultaneous presence of both the modes symmetrically for positive and negative $q$ values. Consequently, unlike the finite-$B_y$ case, no switching between distinct Floquet Majorana sectors occurs across opposite momentum values. This feature is fully consistent with the corresponding self-consistent mean-field results discussed above.



\end{document}